\documentclass[aps,prc,superscriptaddress,showpacs,nofootinbib,twocolumn]{revtex4}

\usepackage[dvips]{graphicx}
\usepackage{amssymb,amsmath,amstext,amsthm,amsfonts}
\usepackage{slashed}
\usepackage{hyperref}
\usepackage{bbm}

\begin{document}


\newcommand{\singleplotscale}{.80}
\newcommand{\doubleplotscale}{.80}

\newcommand{\eb}{\begin{equation}}
\newcommand{\ee}{\end{equation}}
\newcommand{\myunit}[1]{\mbox{$\,\text{#1}$}}
\newcommand{\MeV}{\myunit{MeV}}
\newcommand{\fm}{\myunit{fm}}
\newcommand{\GeV}{\myunit{GeV}}
\newcommand{\GeVsq}{\mbox{$\,\text{GeV}^2$}}
\newcommand{\GeVminsq}{\mbox{$\,\text{GeV}^{-2}$}}
\newcommand{\cmsq}{\mbox{$\,\text{cm}^2$}}
\newcommand{\nucmsq}{\mbox{$\,10^{-38}$\cmsq}}
\newcommand{\myvec}[1]{\boldsymbol{\mathrm{#1}}}        

\newcommand{\fig}[1]{Fig.~#1}
\newcommand{\reffig}[1]{\fig{\ref{#1}}}
\newcommand{\eq}[1]{Eq.~(#1)}
\newcommand{\refeq}[1]{\eq{\ref{#1}}}
\newcommand{\refch}[1]{Chapter~\ref{#1}}
\newcommand{\refsec}[1]{Section~\ref{#1}}
\newcommand{\refapp}[1]{Appendix~\ref{#1}}
\newcommand{\refcite}[1]{Ref.~\cite{#1}}
\newcommand{\refscite}[1]{Refs.~\cite{#1}}
\newcommand{\etal}{\emph{et~al.}}
\newcommand{\refetal}[1]{\etal~\cite{#1}}
\newcommand{\reftab}[1]{Table~\ref{#1}}

\newcommand{\kpi}{k_\pi}
\newcommand{\kpiz}{k_\pi^0}
\newcommand{\slashkpi}{\slashed{k}_\pi}
\newcommand{\kpicn}[1]{k_\pi^{#1}}
\newcommand{\kpico}[1]{{k_\pi}_{#1}}
\newcommand{\abskpi}{|\myvec{\kpi}|}

\newcommand{\kz}{k^0}
\newcommand{\kcn}[1]{k^{#1}}
\newcommand{\kco}[1]{k_{#1}}
\newcommand{\slashk}{\slashed{k}} 
\newcommand{\absk}{|\myvec{k}|}

\newcommand{\kpr}{k'}
\newcommand{\kprz}{{k'}^0}
\newcommand{\kprcn}[1]{k'\,^{#1}}
\newcommand{\kprco}[1]{k'_{#1}}
\newcommand{\slashkpr}{\slashed{k}'} 
\newcommand{\abskpr}{|\myvec{\kpr}|}
\newcommand{\sinth}{\sin \theta}
\newcommand{\costh}{\cos \theta}
\newcommand{\sinph}{\sin \phi}
\newcommand{\cosph}{\cos \phi}

\newcommand{\pz}{p^0}
\newcommand{\pcn}[1]{p^{#1}}
\newcommand{\pco}[1]{p_{#1}}
\newcommand{\slashp}{\slashed{p}} 
\newcommand{\absp}{|\myvec{p}|}

\newcommand{\qz}{q^0}
\newcommand{\slashq}{\slashed{q}} 
\newcommand{\absq}{|\myvec{q}|}

\newcommand{\ppr}{p'}
\newcommand{\pprz}{{p'}^0}
\newcommand{\pprcn}[1]{p'\,^{#1}}
\newcommand{\pprco}[1]{p'_{#1}}
\newcommand{\slashppr}{\slashed{p}'} 
\newcommand{\absppr}{|\myvec{\ppr}|}

\newcommand{\M}{M}
\newcommand{\Mpr}{M'\hspace{.2mm}}
\newcommand{\mpi}{m_\pi}
\newcommand{\mnu}{m_\nu}
\newcommand{\ml}{m_\ell}
\newcommand{\mlpr}{m_{\ell'}}

\newcommand{\tinyMM}[1]{\text{\begin{tiny}#1\end{tiny}}}
\newcommand{\ME}{\mathcal{M}}
\newcommand{\Res}{R}

\newcommand{\subQE}{\text{QE}}
\newcommand{\subhalf}{\text{1/2}}
\newcommand{\subthreehalf}{\text{3/2}}
\newcommand{\subR}{\text{R}}
\newcommand{\subN}{\text{N}}
\newcommand{\subCC}{\text{CC}}
\newcommand{\subNC}{\text{NC}}
\newcommand{\subEM}{\text{EM}}
\newcommand{\subBG}{\text{BG}}
\newcommand{\subtot}{\text{tot}}

\newcommand{\g}[1]{\gamma^{#1}}
\newcommand{\gF}{\gamma^5}
\newcommand{\gN}{\gamma^0}
\newcommand{\gmu}{\gamma^\mu}
\newcommand{\unittwo}{\mathbbm{1}_{2}}
\newcommand{\unitfour}{\mathbbm{1}_{4}}

\newcommand{\Tr}{\mbox{Tr}}
\newcommand{\pr}[1]{{#1}'}
\newcommand{\RealPart}{\text{Re}}
\newcommand{\ImaginaryPart}{\text{Im}}

\newcommand{\ii}{\mathrm{i}}
\newcommand{\dd}{\mathrm{d}}
\newcommand{\dfp}[1]{ \frac{\dd^4 {#1} }{(2\pi)^3} }
\newcommand{\dsdqs}{\frac{\dd \sigma}{\dd Q^2}}
\newcommand{\fff}{\mathcal{F}}
\newcommand{\ffc}{\mathcal{C}}
\newcommand{\ffg}{\mathcal{G}}
\newcommand{\jj}{\mathcal{J}}

\newcommand{\atom}[2]{\mbox{$^{#1}$#2}}
\newcommand{\carbon}{{\atom{12}{C}}}
\newcommand{\oxygen}{{\atom{16}{O}}}
\newcommand{\calcium}{{\atom{40}{Ca}}}
\newcommand{\iron}{{\atom{56}{Fe}}}

\newcommand{\res}[3]{#1$_{#2}$(#3)}
\newcommand{\lep}{\ell}
\newcommand{\coscab}{\mbox{$\cos \theta_C$}}
\newcommand{\sinwein}{\mbox{$\sin^2 \theta_W$}}

\newcommand{\matrixelement}[3]{\left \langle {#1} \left\lvert \vphantom{ {#1} {#2} {#3}} {#2} \right\rvert {#3} \right\rangle}
\newcommand{\clebsch}[2]{\left( {#1} \left\lvert \vphantom{{#1}{#2}} \right. {#2} \right)}
\renewcommand{\binom}[2]{\left\{ \begin{array}{c} #1 \\ #2  \end{array} \right\}}

\newcommand{\lsone}{solid }
\newcommand{\lstwo}{dashed }
\newcommand{\lsthree}{long-dashed }

\newcommand{\eKomma}{\, ,}
\newcommand{\ePunkt}{\, .}
\newcommand{\eSemikolon}{\, ;}

\newcommand{\emailUs}[2]{\footnote[#1]{Electronic address: \texttt{#2}}}

\title{Electron- and neutrino-nucleus scattering from the quasielastic to the resonance region}

\author{T.~Leitner\emailUs{1}{tina.j.leitner@theo.physik.uni-giessen.de}}
\affiliation{Institut f\"ur Theoretische Physik, Universit\"at Giessen, Germany}

\author{O.~Buss\emailUs{3}{oliver.buss@theo.physik.uni-giessen.de}}
\affiliation{Institut f\"ur Theoretische Physik, Universit\"at Giessen, Germany}

\author{L.~Alvarez-Ruso}
\affiliation{Departamento de F\'{\i}sica, Universidad de Murcia, Spain}

\author{U.~Mosel}
\affiliation{Institut f\"ur Theoretische Physik, Universit\"at Giessen, Germany}

\date{\today}

\begin{abstract}
We present a model for electron- and neutrino-scattering off nucleons and nuclei focussing on the quasielastic and resonance region. The lepton-nucleon reaction is described within a relativistic formalism that includes, besides quasielastic scattering, the excitation of 13 $N^*$ and $\Delta$ resonances and a non-resonant single-pion background. Recent electron-scattering data is used for the state-of-the-art parametrizations of the vector form factors; the axial couplings are determined via PCAC and, in the case of the $\Delta$ resonance, the axial form factor is refitted using neutrino-scattering data. Scattering off nuclei is treated within the Giessen Boltzmann-Uehling-Uhlenbeck (GiBUU) framework that takes into account various nuclear effects: the local density approximation for the nuclear ground state; mean-field potentials and in-medium spectral functions. Results for inclusive scattering off Oxygen are presented and, in the case of electron-induced reactions, compared to experimental data and other models.
\end{abstract}

\pacs{25.30.-c}

\maketitle

\section{Introduction}

Scattering electrons off the nucleon, a wealth of information on its properties such as,
e.g., structure functions, charge radius, the distribution of its quark and gluon
constituents and its excitation spectrum has been gathered. Under investigation are also
how these properties change when embedding the nucleon or the resonances in a nuclear
medium. Both the resonances and the nucleon acquire complex self-energies within in the
medium which may lead both to mass-shifts and modifications of the life-times. Calculating
these with a given interaction model and comparing such predictions to the measured
quantities, one is testing our understanding of the hadronic many body problem.  Such
studies have first been performed via inclusive experiments with nuclear targets where
only the outgoing electron was detected (for a recent review cf.,
e.g.,~\refcite{Benhar:2006wy}). Studying semi-inclusive processes in the intermediate
energy regime with an energy transfer of $0.1-2\GeV$, one gains further sensitivity on
in-medium changes of baryonic resonances and the nucleon since such modifications may also
lead to unexpected final state interaction patterns. A related topic of present interest is the
modification of mesons such as, e.g., $\sigma$~\cite{Bloch:2007ka} and
$\omega$~\cite{kotulla:2008xy} within the medium due to chiral symmetry restoration and/or
collisional broadening. A correct understanding of final state interactions is necessary to
distinguish profane (e.g., pion rescattering in the nucleus) from spectacular
(e.g., chiral symmetry restoration) effects.

Closely related is the scattering of neutrinos on nucleons, where current and future
experiments propose to shed light on the nucleon axial form factor and its strange quark
content. However, all those neutrino experiments use nuclear targets and, therefore, all
measured cross sections incorporate nuclear effects~\cite{nuint}. To draw conclusions on
the underlying $\nu N$ process it is therefore necessary to understand these corrections.
Nowadays, the interest in neutrino nucleus reactions is driven by the discovery of
neutrino oscillations---there is an extensive experimental effort aiming at a precise
determination of neutrino oscillation parameters as mixing angles, neutrino mass squared
differences and possible CP violation \cite{k2k,miniboone,sciboone,t2k,minos,minervaprop}.
This demands for an equally precise knowledge of the neutrino nucleus interaction process.
A critical quantity is the neutrino energy which can not be measured directly but has to
be reconstructed from observables, i.e., in the case of charged current (CC) reactions from
the outgoing charged lepton, or in the case of neutral currents (NC) from the hadronic debris leaving the nucleus.
Neutrino induced pion production is strongly influenced by nuclear effects,
its understanding is crucial since
NC $\pi^0$ production is a major background in $\nu_e$ appearance experiments,
while CC $\pi^+$ production introduces a background to $\nu_\mu$ disappearance searches.
A good knowledge of neutrino-nucleus interactions is
thus necessary to minimize the systematic uncertainties in neutrino fluxes and
backgrounds.

Electron and neutrino scattering off the nucleon are closely interconnected and one can treat
both processes within the same formalism. Furthermore, the proper description of the electron
nucleus interaction serves as a necessary benchmark for the nuclear corrections in the neutrino
nucleus reaction.

Many authors investigated electron \emph{or} neutrino scattering on nucleons, but only a
few study both within the same model.  In the intermediate energy regime, the
lepton-nucleon cross section is dominated by elastic scattering and pion production via
resonance excitations or non-resonant processes.  One approach to describe pion production
is based on helicity amplitudes derived in a relativistic quark model
\cite{Feynman:1971wr}. Those have been adopted to describe both, electro- and
neutrino production (via CC and NC) of resonances \cite{Ravndal:1972pn, Rein:1980wg}.
Nieves and collaborators extended their model for $e N \to e' N' \pi$ \cite{Gil:1997bm} to
describe CC and NC pion production on the nucleon \cite{Hernandez:2007qq}. Besides the
dominant $\Delta$ contribution they included background terms required by chiral symmetry.
Also the model of Sato and Lee for neutrino induced pion production on the nucleon
including $\Delta$ excitation and non-resonant background terms \cite{Sato:2003rq} is
based on their approach for electrons \cite{Sato:2000jf}.

The influence of nuclear effects on both, electron and neutrino scattering cross sections,
has been investigated by Benhar \emph{et al.} using an impulse approximation model with
realistic spectral functions obtained from nuclear many body theory calculations
\cite{Benhar:2005dj,Nakamura:2007pj}. In particular in the quasielastic peak region, they
achieve good agreement to inclusive electron scattering data.
Nuclear effects in the QE
region have also been investigated in detail by Nieves {\etal} for electrons
\cite{Gil:1997bm} and neutrinos \cite{Nieves:2004wx,Nieves:2005rq} where they have
included, among other nuclear corrections, long range nuclear correlations. Also this
approach describes inclusive electron scattering data with impressive agreement.
A relativistic Green's function approach was applied by Meucci~\refetal{Meucci:2003uy,Meucci:2003cv}
to inclusive electron as well as to inclusive neutrino nucleus reactions.
Butkevich \refetal{Butkevich:2005ph,Butkevich:2007gm} addresses both neutrino and electron
scattering with special emphasis on the impact of different impulse approximation (IA)
schemes: plane wave IA (PWIA) and relativistic distorted wave IA (RDWIA). Also the Ghent group apply
RPWIA and RDWIA models to neutrino and electron scattering in the QE region~\cite{Martinez:2005xe};
lately they extended their framework to pion production~\cite{Praet:2008yn}.

In this article, we concentrate on two issues: First, we consider electron and neutrino
--- both charged (CC) and neutral current (NC) --- scattering off the nucleon. Then, we
investigate the influence of the nuclear medium on inclusive electron and neutrino cross
sections. Within our Giessen BUU~(GiBUU) framework, we aim at a consistent treatment of
the initial vertex and the final state processes and we emphasize that these should not be
treated separately.

Compared to our previous work \cite{Buss:2007ar}, where only the $\Delta$ resonance was
included, we have improved our model considerably: Besides the
$\Delta$ we have included 12 higher resonances. Recent electron scattering data were used
to fix the resonance vector form factors; but also the axial ones were refitted. A
non-resonant background is accounted for also in the neutrino case. Semi-inclusive
processes such as pion production or nucleon knockout will be discussed in a forthcoming
publication and are not considered in this work. First results, where we included only QE and
$\Delta$ excitation, have already been presented in \refscite{Leitner:2006ww,Leitner:2006sp} for
CC and NC neutrino induced reactions.

This article is organized as follows. First, we present our model for the electron and
neutrino nucleon interaction where we put the emphasis on the neutrino-nucleon reaction.
Then, we discuss the nuclear ground state and the modifications of the elementary
lepton-nucleon vertex inside the nucleus. Afterwards, results for both electron and
neutrino inclusive scattering off Oxygen are presented. Main attention is given to the
electron-nucleus reaction where experimental data allow for conclusive comparisons. The
effect of different nuclear corrections is discussed.  Finally, we compare our model to other approaches.
The appendices complete Sec.~II by
giving all necessary details.

\section{Lepton-nucleon interaction}
\label{sec:leptonnucleon}

In this section, we explain in detail our model for the lepton-nucleon interaction. In the region of
intermediate lepton beam energies ($E_{beam}\sim 0.5-2\GeV$), the cross section contains
contributions from quasielastic (QE) scattering ($ \ell N \to \ell' N'$), resonance (R)
excitation ($ \ell N \to \ell' R$) and direct, i.e.~non-resonant, single-pion production ($ \ell N \to \ell'
\pi N'$) treated in our description as background (BG). Thus we assume
\begin{equation}
  \dd \sigma_\subtot=\dd \sigma_\subQE + \sum_R \dd \sigma_\subR + \dd \sigma_\subBG \ePunkt  \label{eq:generalDecomposition}
\end{equation}

The dynamics of the interaction is encoded in the absolute value of the matrix element squared, summed and averaged over initial and final spins,
\begin{equation}
  |\bar{\ME}_{\subQE,\subR,\subBG}|^2 = C_{\subEM,\subCC,\subNC}^2 L_{\mu \nu} H^{\mu \nu}_{\subQE,\subR,\subBG} \eKomma
\end{equation}
with $C_{\subEM}=4 \pi \alpha/q^2$, $C_{\subCC}=G_F \cos \theta_C / \sqrt{2}$ or $C_{\subNC}= G_F /\sqrt{2}$. $Q^2$ is the four-momentum transfer; $\alpha=1/137$, $G_F = 1.16637 \cdot 10^{-5}\GeVminsq $ and $ \coscab = 0.9745 $.
The leptonic tensor is given by
\begin{align}
  L_{\mu \nu}= \frac{1+a}{2} \Tr \left[ (\slashk + m_{\ell} ) \tilde{l}_\mu   (\slashkpr + m_{\ell'} ) l_\nu \right] \eKomma \label{eq:leptontensor}
\end{align}
where $l_\mu=\gamma_{\mu} (1- a \gamma_5)$ and $\tilde{l}_\mu=\gamma_0 l_\mu^\dagger \gamma_0$; $k$ ($\kpr$) denotes the 4-vector of the incoming (outgoing) lepton and $m_\ell$ ($m_{\ell'}$) the corresponding masses.  The parameter
$a$ depends on the reaction process: $a=0$ for EM and $a=1$ for CC or NC neutrino
scattering.

The hadronic currents in
$H^{\mu \nu}_{\subQE,\subR,\subBG}$ have to be parametrized in terms of form factors and thus depend not only on
the final state but also on the specific process, namely electromagnetic (EM) ($e^- N
\rightarrow e^- X $), charged current (CC) ($\nu_\ell N \rightarrow \ell^- X$) or neutral
current (NC) ($\nu N \rightarrow \nu X$).


\subsection{Quasielastic scattering}
\label{ch:quasielastic}

The cross section for quasielastic scattering $\ell(k) N(p) \to \ell'(\kpr) N'(\ppr)$ is
given by\footnote{Throughout this article we work in the lab frame if not noted otherwise.}
\begin{align}
  \frac{\dd\sigma_{\subQE}}{\dd \omega \; \dd \Omega_{\kpr}} = \frac{\abskpr}{32 \pi^2} \;
  \frac{ \delta (\ppr^2 - \Mpr^2) }{\left[ (k \cdot p)^2- \ml^2 M^2\right]^{1/2}} \; |
  \bar{\ME_{\subQE}} |^2 \eSemikolon \label{eq:QEdoublediff_xsec}
\end{align}
with $M= \sqrt{p^2}$ and $\Mpr= \sqrt{{\ppr}^2}$. In the case of \emph{free nucleons}, we have
$M=\Mpr=M_N$, where $M_N$ denotes the average nucleon mass for which we take 938\MeV; $\omega=\kz-\kprz$ is the energy transfer and $\Omega_{\kpr}$ is the solid angle between incoming and outgoing leptons.

The hadronic tensor $H^{\mu \nu}_{\subQE}$ for quasielastic scattering  is given by
\begin{align}
  H^{\mu \nu}_{\subQE} = \frac12 \Tr \left[ (\slashp + M ) \tilde{\Gamma}^{\mu}_{\subQE}
    (\slashppr + \Mpr ) \Gamma^{\nu}_{\subQE} \right] \eKomma \label{eq:QEhadronictensor}
\end{align}
with
\begin{equation}
  \tilde{\Gamma}^{\mu}_{\subQE}=\gamma_0 {\Gamma^{\mu}_{\subQE}}^{\!\!\!\!\dagger} \gamma_0 \ePunkt
\end{equation}
$\Gamma^{\mu}_{\subQE}$ has a $V-A$ Lorentz structure
\begin{equation}
  \Gamma^{\mu}_{\subQE}=\mathcal{V}^{\mu}_{\subQE} - \mathcal{A}^{\mu}_{\subQE} \eKomma \label{eq:QEcurrent}
\end{equation}
with the vector part
\begin{equation}
  \mathcal{V}^{\mu}_{\subQE}=\fff_1 \gamma^\mu  + \frac{\fff_2}{2 M_N} \ii \sigma^{\mu\alpha} q_\alpha \eKomma
   \label{eq:QEVECcurrent}
\end{equation}
and the axial part
\begin{equation}
  -\mathcal{A}^{\mu}_{\subQE}= \fff_A \gamma^\mu \gamma_5 + \frac{\fff_P}{M_N} q^\mu \gamma_5 \ePunkt
 \label{eq:QEAXcurrent}
\end{equation}
Here, $q_{\mu}=p'_{\mu}-p_{\mu}$.
$\fff_i$ ($i=1,2$) stands either for the CC form factors $F_i^V$, for the NC form factors $\tilde{F}_i^N$ or the EM form factors $F^N_i$ with $N=p,n$; $\fff_A$ for the CC form factor $F_A$ and the NC form factor $\tilde{F}_A^N$ (analogous for $\fff_P$). All form factors depend on $Q^2=-q^2$.

The vector form factors $F_i^V$ and $\tilde{F}_i^N$ can be related to the electromagnetic Dirac and Pauli
form factors $F^N_i$ with $N=p,n$ as listed in \reftab{tab:ff}. Those can be rewritten
in terms of Sachs form factors, for which we take the updated BBBA-2007 parametrization
\cite{Bodek:2007vi}. The relations for the axial form factors are also given in
\reftab{tab:ff}. In the cross section formula, $\fff_P$ appears only multiplied by the mass
of the outgoing lepton, so it can be safely ignored for NC interactions. In the case of CC
reactions, we assume pion pole dominance, and use the partial conservation of the axial current (PCAC) to relate $F_A$ and $F_P$ (details are given in Appendix
\ref{sec:iso12iso12}),
\begin{equation}
  F_P(Q^2)=\frac{2 M_N^2}{Q^2+m_{\pi}^2}F_A(Q^2) \ePunkt \label{eq:pseudoff}
\end{equation}
For the axial form factor, we assume a standard dipole form
\begin{equation}
  F_A(Q^2)=g_A \left(1+\frac{Q^2}{M_A^2}\right)^{-2}\eSemikolon  \label{eq:axialff}
\end{equation}
with the axial vector coupling constant $g_A=-1.267$ \cite{Yao:2006px}.  The so-called axial mass
$M_A$ has recently been refitted by Kuzmin \refetal{Kuzmin:2007kr} using the BBBA-2007 parametrization for the
vector form factors. We take their best fit value
\begin{equation}
  M_A=0.999 \pm 0.011 \GeV \ePunkt  \label{eq:MA}
\end{equation}

The strangeness content of the nucleon is contained in $F_{1,2}^{s}$ and $F_{A}^{s}$ (see
\reftab{tab:ff}) which are present in NC neutrino scattering. No definite conclusion can
be drawn from data, thus we set for simplicity
\begin{align}
  F_1^{s}(0)&= 0 \eKomma \\
  F_2^{s}(0)&= 0 \eKomma
\end{align}
and
\begin{equation}
  F_A^{s}(Q^2)=\frac{\Delta s}{\left(1+\frac{Q^2}{M_A^2}\right)^2} \eKomma
\end{equation}
assuming that the strange axial mass is equal to the non-strange one. Here $\Delta s$
denotes the strangeness contribution to the nucleon spin for which we use $\Delta s= - 0.15$.

%
\begin{table*}[t]
  \begin{center}
    \vspace{1cm}
    \begin{tabular}{ c c c c c c}
      \noalign{\vspace{-8pt}}
      \hline \hline
      & \multicolumn{2}{c}{replace $\fff_i$ with} &$\quad$ &  \multicolumn{2}{c}{replace $\fff_A$ with}  \\  
      reaction   & for  $I=1/2$ &  for $I=3/2$ & & for  $I=1/2$ &  for $I=3/2$  \\
      \hline
      $e^- p \rightarrow e^- {X}^+$       &  $F^p_i$       & $F^N_i$ & &    -       & - \\
      $e^- n \rightarrow e^- {X}^0$       &  $F^n_i$       & $F^N_i$ & &    -       & - \\
      $\nu p \rightarrow \ell^- {X}^{++}$ &  -             & $F_i^V=-\sqrt{3} F^N_i$ & &    -       & $\sqrt{3} F_A$ \\
      $\nu n \rightarrow \ell^- {X}^{+}$  &  $F_i^V=F^p_i-F^n_i$ & $F_i^V=-F^N_i$  & &    $F_A$   & $F_A$ \\
      $\nu p \rightarrow \nu {X}^{+}$     &  $\tilde{F}_i^p=(\frac12 -2 \sinwein) F_i^p - \frac12 F_i^n - \frac12 F_i^s$ & $\tilde{F}_i^N=-(1-2 \sinwein) F^N_i$ && $\tilde{F}_A^p=\frac12 F_A + \frac12 F_A^s$ & $\tilde{F}_A^N=F_A$ \\
      $\nu n \rightarrow \nu {X}^{0} $  &     $\quad \tilde{F}_i^n=(\frac12 -2 \sinwein) F_i^n - \frac12 F_i^p - \frac12 F_i^s \quad $ & $\tilde{F}_i^N=-(1-2 \sinwein) F^N_i $ &&  $\quad \tilde{F}_A^n=-\frac12 F_A + \frac12 F_A^s \quad$ & $ \tilde{F}_A^N=F_A $ \\
      \hline \hline
    \end{tabular}
  \end{center}
  \caption{Isospin relations for the form factors. $X$ stands for the nucleon $N$, and $N^*$ or $\Delta$ resonances. $\fff_i$ is the generalized vector form factor in \refeq{eq:QEVECcurrent} and \refeq{eq:vectorspinhalfcurrent}, which has --- depending on the process --- to be substituted following the prescription in the table (analogous for $\fff_A$ in \refeq{eq:QEAXcurrent} and \refeq{eq:axialspinhalfcurrent}). Note that in the case of an isospin 1/2 $\to$ 3/2 transition, the form factors are equal for proton and neutron which is indicated by the index $N$ (instead of $p$ or $n$). \textbf{This replacement scheme is identical for both, spin 1/2 and spin 3/2 resonance excitations. In the case of spin 3/2, replace $\fff$ with $\ffc$ (cf.~\refeq{eq:vectorspinthreehalfcurrent} and \refeq{eq:axialspinthreehalfcurrent}), and $F$ with $C$.}
 A detailed derivation of these relations is given in Appendix \ref{sec:iso12iso12} for $I=1/2$ resonances and in Appendix \ref{sec:iso12iso32} for $I=3/2$ resonances. \label{tab:ff}}
\end{table*}

\subsection{Excitation of baryon resonances}
\label{ch:excitationofResonances}

This section is devoted to the second part in our general decomposition of the cross
section given in \refeq{eq:generalDecomposition}, namely the excitation of resonances
$\sum_R \dd \sigma_R$. As will be shown, the electromagnetic form
factors are taken from the MAID analysis \cite{MAIDWebsite,Tiator:2006dq,Drechsel:2007if}. In this analysis, 13 resonances with invariant masses of less than $2$ GeV
are included. This limits then also the number of resonances in our model to 13 --- they are listed in \reftab{tab:included_resonances}.

The cross section for resonance excitation $\ell(k) N(p) \to \ell'(\kpr) R(\ppr)$ is given
by
\begin{align}
  \frac{\dd\sigma_{\subR}}{\dd \omega \; \dd \Omega_{\kpr}} = \frac{\abskpr}{32 \pi^2} \;
  \frac{ \mathcal{A}(\ppr^2) }{\left[ (k \cdot p)^2- \ml^2 M^2\right]^{1/2}} \; |
  \bar{\ME_\subR} |^2 \eSemikolon \label{eq:RESdoublediff_xsec}
\end{align}
with $M= \sqrt{p^2}$.  $\mathcal{A}(\ppr^2)$ denotes the vacuum spectral function of the
particle, which is given by a Breit-Wigner distribution
\begin{align}
  \mathcal{A}(\ppr^2) = \frac{\sqrt{\ppr^2}}{\pi} \frac{\Gamma(\ppr)}{(\ppr^2-M_R^2)^2 + \ppr^2 \Gamma^2(\ppr)} \eKomma
\end{align}
with the momentum-dependent width $\Gamma$ taken from the Manley analysis \cite{manley} (see
\reftab{tab:included_resonances} for the values at the pole).

\begin{table*}[t]
  \begin{center}
    \vspace{1cm}
    \begin{tabular}{c c c c c c c c}
      \noalign{\vspace{-8pt}}
      \hline \hline
      name               & $\quad  M_R$ [GeV]$ \quad$ & $\quad  J \quad $   & $ \quad I\quad  $   & $\quad  P \quad $ & $\quad \Gamma_0^\subtot$ [GeV]$\quad $ &  $\quad \pi N$ branching ratio $\quad $   & $\quad F_A(0)$ or  $C_5^A(0)$ \\ \hline
      \res{P}{33}{1232}  &       $1.232$ & $3/2$ &$ 3/2$ &$ +$ &                $ 0.118$ &                              $ 1.00$ &$\phantom{-} 1.17$   \\
      \res{P}{11}{1440}  &       $1.462$ & $1/2$ &$ 1/2$ &$ +$ &                $ 0.391$ &                              $ 0.69$ &$ -0.52$\\
      \res{D}{13}{1520}  &       $1.524$ & $3/2$ &$ 1/2$ &$ -$ &                $ 0.124$ &                              $ 0.59$ &$-2.15$\\
      \res{S}{11}{1535}  &       $1.534$ & $1/2$ &$ 1/2$ &$ -$ &                $ 0.151$ &                              $ 0.51$ &$-0.23$\\
      \res{S}{31}{1620}  &       $1.672$ & $1/2$ &$ 3/2$ &$ -$ &                $ 0.154$ &                              $ 0.09$ &$\phantom{-}0.05$\\
      \res{S}{11}{1650}  &       $1.659$ & $1/2$ &$ 1/2$ &$ -$ &                $ 0.173$ &                              $ 0.89$ &$-0.25$\\
      \res{D}{15}{1675}  &       $1.676$ & $5/2$ &$ 1/2$ &$ -$ &                $ 0.159$ &                              $ 0.47$ &$-1.38$\\
      \res{F}{15}{1680}  &       $1.684$ & $5/2$ &$ 1/2$ &$ +$ &                $ 0.139$ &                              $ 0.70$ &$ -0.43$\\
      \res{D}{33}{1700}  &       $1.762$ & $3/2$ &$ 3/2$ &$ -$ &                $ 0.599$ &                              $ 0.14$ &$\phantom{-} 0.84$\\
      \res{P}{13}{1720}  &       $1.717$ & $3/2$ &$ 1/2$ &$ +$ &                $ 0.383$ &                              $ 0.13$ &$ -0.29$\\
      \res{F}{35}{1905}  &       $1.881$ & $5/2$ &$ 3/2$ &$ +$ &                $ 0.327$ &                              $ 0.12$ &$ \phantom{-}0.15$\\
      \res{P}{31}{1910}  &       $1.882$ & $1/2$ &$ 3/2$ &$ +$ &                $ 0.239$ &                              $ 0.23$ &$\phantom{-}0.08$\\
      \res{F}{37}{1950}  &       $1.945$ & $7/2$ &$ 3/2$ &$ +$ &                $ 0.300$ &                              $ 0.38$ &$\phantom{-} 0.24$\\
      \hline \hline
    \end{tabular}
  \end{center}
  \caption{Properties of the resonances included in our model. The Breit-Wigner mass $M_R$, spin $J$, isospin $I$, parity $P$, the vacuum total decay width $\Gamma_0^\subtot$ at the pole, the branching ratio into $\pi N$ and the axial coupling ($F_A(0)$ for spin 1/2 states; $C_5^A(0)$ for states with spin 3/2 or higher) are listed. The resonance parameters are taken from the analysis of Manley \refetal{manley}.}
   \label{tab:included_resonances}
\end{table*}

\subsubsection{Excitation of spin 1/2 resonances}
\label{sec:excspinhalf}

The hadronic tensor for the excitation of a spin 1/2 resonance is given by
\begin{align}
  H^{\mu \nu}_{1/2} = \frac12 \Tr \left[ (\slashp + M ) \tilde{\Gamma}^{\mu}_{1/2}
    (\slashppr + \Mpr ) \Gamma^{\nu}_{1/2} \right]\eKomma
\end{align}
with $M= \sqrt{p^2}$ and $\Mpr= \sqrt{{\ppr}^2}$ and
\begin{equation}
  \tilde{\Gamma}^{\mu}_{1/2}=\gamma_0 {\Gamma^{\mu}_{1/2}}^{\!\!\!\!\dagger} \gN.
\end{equation}
For states with positive parity (e.g.~\res{P}{11}{1440}),
\begin{equation}
  \Gamma^{\mu}_{1/2+}=\mathcal{V}^{\mu}_{1/2} - \mathcal{A}^{\mu}_{1/2}
\end{equation}
and for states with negative parity (e.g.~\res{S}{11}{1535}),
\begin{equation}
  \Gamma^{\mu}_{1/2-}=\left[ \mathcal{V}^{\mu}_{1/2} - \mathcal{A}^{\mu}_{1/2} \right] \gF \eKomma \label{eq:spinhalfcurrentnegparity}
\end{equation}
where the vector part $\mathcal{V}^{\mu}_{1/2}$ is given by
\begin{equation}
  \mathcal{V}^{\mu}_{1/2}=\frac{\fff_1}{(2 M_N)^2} \left( Q^2 \gamma^\mu + \slashq q^\mu \right) + \frac{\fff_2}{2 M_N} \ii \sigma^{\mu\alpha} q_\alpha  \label{eq:vectorspinhalfcurrent}
\end{equation}
and the axial part $\mathcal{A}^{\mu}_{1/2}$ by
\begin{equation}
  -\mathcal{A}^{\mu}_{1/2}= \fff_A \gamma^\mu \gamma_5 + \frac{\fff_P}{M_N} q^\mu \gF \ePunkt \label{eq:axialspinhalfcurrent}
\end{equation}
As in the QE case, $\fff_i$ ($i=1,2$) stands either for the CC form factors $F_i^V$, for the NC form factors $\tilde{F}_i^N$ or the EM form factors $F^N_i$ with $N=p,n$; analogous for $\fff_A$ and $\fff_P$. The form factors depend on $Q^2=-q^2$.

$F_i^V$ can be related to the electromagnetic transition form factors $F^N_i$ with
$N=p,n$ as listed in \reftab{tab:ff}, thus the form factors for neutrino and electron
scattering are related.
The form factors $F^{p,n}_i$ can be derived from helicity amplitudes extracted from
electron scattering experiments. The explicit relations between the form factors
$F_i^{p,n}$ and the helicity amplitudes $A_{1/2}^{p,n}$, and $S_{1/2}^{p,n}$ are given in
Appendix \ref{sec:heliFF_1_2} for both, positive and negative parity states. We use
these relations to extract the form factors from the results of the recent MAID2005
analysis \cite{tiatorprivcomm,MAIDWebsite,Tiator:2006dq,Drechsel:2007if} for
the helicity amplitudes and their $Q^2$-dependence.

Experimental information on the $N-R$ axial form factors $F_A$ and $F_P$ is very limited.
Goldberger-Treiman relations have been derived for the axial couplings \cite{Fogli:1979cz}, but there is no information about the $Q^2$-dependence. We will follow
this approach and apply PCAC and pion pole dominance to derive the axial
couplings $F_A(0)$ and to relate $F_A$ and $F_P$. The derivation, performed in
Appendix~\ref{sec:spinhalfpcac}, leads to
\begin{equation}
  F_P(Q^2)= \frac{(M_R \pm M_N)M_N}{Q^2+m_{\pi}^2}F_A(Q^2) \eKomma
\end{equation}
with $+$ ($-$) for positive (negative) parity resonances.  The $Q^2$-dependence of the
axial form factor is neither fully constrained by theory nor by experiment, so we assume --- as
in the QE case --- a dipole dependence
\begin{equation}
  F_A(Q^2)=F_A(0) \left(1+\frac{Q^2}{{M_A^*}^2}\right)^{-2} \eSemikolon \label{eq:axialspinhalfff}
\end{equation}
with $M_A^*=1\GeV$ as for the nucleon.  The coupling $F_A(0)$ can be related to the $\pi N R$-coupling as
detailed in Appendix~\ref{sec:spinhalfpcac} (non-diagonal Goldberger-Treiman relation) and
we obtain the results summarized in \reftab{tab:included_resonances}.

As in the nucleon case, strange form factors can
contribute for isospin 1/2 $\to$ 1/2 transitions (cf.~Appendix \ref{sec:iso12iso12}). However, the present
experimental status does not allow any conclusions on the strange transition form factors,
thus, we neglect them in this work and set $F_i^s$ and $F_A^s$ to zero.

Several studies have addressed the properties of the \res{P}{11}{1440} as well as its electromagnetic and weak production~\cite{Li:1991yb, Cano:1998wz, Alberto:2001fy, Cardarelli:1996vn, Alvarez-Ruso:1997jr, Dong:1999cz, Alvarez-Ruso:2003gj, Hernandez:2007ej}.
In the works of \refscite{Fogli:1979cz,Lalakulich:2006sw} the form factors of the four lowest lying resonances, \res{P}{33}{1232}, \res{P}{11}{1440}, \res{D}{13}{1520} and \res{S}{11}{1535}, are extracted using helicity amplitudes for the vector form factors and PCAC for the determination of the axial couplings.

We now compare our form factors to the most recent works by Lalakulich \refetal{Lalakulich:2006sw} and Hernandez \refetal{Hernandez:2007ej}. To perform a meaningful comparison, we have corrected for different normalizations\footnote{which, in both cases, introduces factors $\frac{(2 M_N)^2}{\mu^2}$ $\left(\frac{2 M_N}{\mu}\right)$ with $\mu=M_N+M_R$ in front of the vector form factors} and sign conventions: both use a different sign in the definition of the axial current.
The result after these corrections is shown in \reftab{tab:compspinhalf_vecff}.
We agree in all signs with Hernandez \emph{et al.}, but with little differences in the numerical values. For the vector form factors, these differences can be attributed to the different data set which is used, since their Eqs.~(28) and (29) coincide with our Eqs.~\eqref{eq:spin12heliposA12} and \eqref{eq:spin12heliposS12}\footnote{While we include the minus sign in the $S_{1/2}$ amplitude explicitly, they account for it when comparing to the MAID analysis.}. We also agree in the expressions for the Goldberger-Treiman relation (their Eq.~(21) vs.~our \refeq{eq:spin12_axialcoupling} --- note the different sign convention); the difference in the axial coupling comes from a different value for the $\pi N R$-coupling.

Comparing to Lalakulich \emph{et al.}, we find major differences. Ignoring a global sign difference, a relative sign difference between $F_1^V$ and $F_A$ remains for the \res{P}{11}{1440} resonance. This is mainly caused by the different sign of $F_1^p$, which can be attributed --- as already pointed out by Hernandez \refetal{Hernandez:2007ej} --- to the extra minus sign in the $S_{1/2}$ amplitude which is missing in the work of Lalakulich \emph{et al.}\footnote{If we also do not consider this minus sign, we find reasonable agreement.}. Furthermore, they perform the calculation in the lab frame, while it should be done in the cm frame \cite{tiatorprivcomm,Aznauryan:2008us} (this affects $S_{1/2}$, see also the remarks in \refapp{ch:formfactors_helicityamplitudes}).
\begin{table}[t]
\begin{center}
\vspace{1cm}
\begin{tabular}{ c c c c c}
\noalign{\vspace{-8pt}}
\hline \hline
 & & \hspace{2pt} this work \hspace{2pt} &  Hernandez  &\hspace{2pt} Lalakulich \\
\hline
\res{P}{11}{1440} & $F^p_1(0)$ & $-1.96 $ & $ -3.55 $ & $ \phantom{-}1.43 $  \\
                  & $F^n_1(0)$ & $\phantom{-}2.26  $ & $ \phantom{-}0.26 $ & $ - 1.43$   \\
                  & $F^V_1(0)$ & $-4.22  $ & $ -3.81 $ & $ - 2.86$  \\
                  & $F^p_2(0)$ & $-0.46 $ & $ -0.50 $ & $ -0.60 $  \\
                  & $F^n_2(0)$ & $\phantom{-}0.41  $ & $ \phantom{-}0.34  $ & $ \phantom{-}0.60 $ \\
                  & $F^V_2(0)$ & $-0.87  $ & $ -0.84  $ & $ \phantom{-}1.20 $ \\
                  & $F_A(0) $  & $-0.52 $ & $ -0.63 $ & $  \phantom{-}0.51 $  \\
$\quad$   \\
\res{S}{11}{1535} & $F^p_1(0)$ & $\phantom{-}0.85 $ & $ - $ & $ -1.24 $  \\
                  & $F^n_1(0)$ & $-0.02$ & $ - $ & $ \phantom{-}1.24 $   \\
                  & $F^V_1(0)$ & $\phantom{-}0.87$ & $ - $ & $ -2.49 $   \\
                  & $F^p_2(0)$ & $\phantom{-}0.46 $ & $ - $ & $ -0.66$  \\
                  & $F^n_2(0)$ & $-0.35 $ & $ - $ & $ \phantom{-}0.66$  \\
                  & $F^V_2(0)$ & $\phantom{-}0.82 $ & $ - $ & $ -1.32$  \\
                  & $F_A(0)$   & $-0.23 $ & $ - $ & $ \phantom{-}0.21 $  \\
\hline \hline
\end{tabular}
\end{center}
\caption{Comparison of our form factors for the \res{P}{11}{1440} and the \res{S}{11}{1535} to the ones compiled by Hernandez \refetal{Hernandez:2007ej} and Lalakulich \refetal{Lalakulich:2006sw} taken at $Q^2=0$. The \res{S}{11}{1535} resonance is not considered in the work of Hernandez. We have corrected for different normalizations and sign conventions for the axial current.
 \label{tab:compspinhalf_vecff}}
\end{table}

\subsubsection{Excitation of spin 3/2 resonances}
\label{sec:excitation_spin_threehalf}
The excitation of a spin 3/2 final state is described within a Rarita-Schwinger formalism
where the hadronic tensor is given by
\begin{align}
  H^{\mu \nu}_{3/2}= \frac12 \Tr \left[ (\slashp + M ) \tilde{\Gamma}^{\alpha \mu}_{3/2}
    \Lambda_{\alpha \beta} \Gamma^{\beta \nu}_{3/2} \right] \eKomma
\end{align}
with the spin 3/2 projector
\begin{align}
  \Lambda_{\alpha \beta}=& - \left(\slashppr + \Mpr \right) \nonumber \\ &\times  \left(
    g_{\alpha \beta} - \frac{2}{3} \frac{\ppr_{\alpha } \ppr_{\beta }}{\Mpr^2}
 + \frac{1}{3}
    \frac{\ppr_{\alpha } \gamma_{\beta} - \ppr_{\beta } \gamma_{\alpha}}{\Mpr} - \frac{1}{3}
    \gamma_{\alpha} \gamma_{\beta} \right)\eKomma
\end{align}
and
\begin{equation}
  \tilde{\Gamma}^{\alpha \mu}_{3/2}=\gamma_0 {\Gamma^{\alpha \mu}_{3/2}}^{\dagger} \gamma_0 \ePunkt
\end{equation}

For states with positive parity as the \res{P}{33}{1232}, we have
\begin{equation}
  \Gamma^{\alpha \mu }_{3/2+} = \left[ \mathcal{V}^{\alpha \mu }_{3/2} - \mathcal{A}^{\alpha \mu }_{3/2}\right] \gamma_{5} \eKomma
\end{equation}
and for the negative parity ones (e.g.~\res{D}{13}{1535}),
\begin{equation}
  \Gamma^{\alpha \mu }_{3/2-} = \mathcal{V}^{\alpha \mu }_{3/2} - \mathcal{A}^{\alpha \mu }_{3/2} \ePunkt
\end{equation}
In terms of form factors, the vector part is given by
  \begin{align}
    \mathcal{V}^{\alpha \mu }_{3/2} =&
    \frac{\ffc_3^V}{M_N} (g^{\alpha \mu} \slashq - q^{\alpha} \gamma^{\mu})+
    \frac{\ffc_4^V}{M_N^2} (g^{\alpha \mu} q\cdot \ppr - q^{\alpha} {\ppr}^{\mu}) \nonumber \\
 & + \frac{\ffc_5^V}{M_N^2} (g^{\alpha \mu} q\cdot p - q^{\alpha} p^{\mu}) + g^{\alpha \mu} \ffc_6^V \label{eq:vectorspinthreehalfcurrent}
\end{align}
  and the axial part by
  \begin{align}
    -\mathcal{A}^{\alpha \mu }_{3/2} =& \left[\frac{\ffc_3^A}{M_N} (g^{\alpha \mu} \slashq - q^{\alpha} \gamma^{\mu})+
      \frac{\ffc_4^A}{M_N^2} (g^{\alpha \mu} q\cdot \ppr - q^{\alpha} {\ppr}^{\mu})  \right. \nonumber \\ & \left.
 +      {\ffc_5^A} g^{\alpha \mu}  + \frac{\ffc_6^A}{M_N^2} q^{\alpha} q^{\mu}\right] \gamma_{5} \ePunkt \label{eq:axialspinthreehalfcurrent}
\end{align}
As before, the calligraphic $\ffc$ stands either for the CC form factors $C_i^{V,A}$, $i=3,\ldots,6$, the electromagnetic
transition form factors $C^N_i$ with $N=p,n$ or the NC form factors $\tilde{C}_i^{V,A \; N}$ as detailed in \reftab{tab:ff}. Note that current conservation implies $C_6^N=0$ for EM transitions.

Using information from electron scattering, the $C^{p,n}_i$ can be parametrized in the same way as in the previous section for spin 1/2 resonances by relating them to the MAID helicity amplitudes. The explicit relations
between the form factors $C_i^{p,n}$ and the helicity amplitudes $A_{1/2}^{p,n}$,
$A_{3/2}^{p,n}$ and $S_{1/2}^{p,n}$ are given in Appendix \ref{sec:heliFF_3_2}.
Note that in the case of isovector transitions, i.e., isospin 1/2 $\to$ 3/2 transitions, proton and neutron form factors are identical.

The substitutions for the axial form factors are summarized in \reftab{tab:ff}. Pion pole dominance and PCAC allow us on one side to relate $C_5^A$ and $C_6^A$, and on the other side to fix the coupling $C_5^A(Q^2=0)$. In Appendix
\ref{sec:spinthreehalfpcac} we show that
\begin{equation}
  C_6^A (Q^2)= \frac{M_N^2}{Q^2+m_{\pi}^2} C_5^A(Q^2)
\end{equation}
for both parity states. The $C_6^A$ form factor appears in the cross section only
multiplied by the mass of the outgoing lepton, such that its contribution is negligible in NC and rather small even
in CC reactions (except for $\nu_\tau$).

For isospin 1/2 $\to$ 1/2 transitions, strange form factors can contribute, but are neglected in this work due to the lack of experimental information.

\textbf{\res{P}{33}{1232}.}
The axial coupling $C_5^A(0)$ is obtained using an
off-diagonal Goldberger-Treiman relation (see Appendix~\ref{sec:spinhalfpcac}), which yields the values given in \reftab{tab:included_resonances}.  Since one can not constrain $C_3^{A}(Q^2), C_4^{A}(Q^2)$ and $C_5^{A}(Q^2)/C_5^{A}(0)$ from theory, these form factors have to be extracted from experiment. The available information for such reactions comes mainly from two bubble chamber experiments: the 12-foot bubble chamber at Argonne (ANL)~\cite{Barish:1978pj,Radecky:1981fn} and the 7-foot bubble chamber at Brookhaven
(BNL)~\cite{Kitagaki:1986ct,Kitagaki:1990vs}.

\begin{figure}[tbp]
  \centerline{\includegraphics[scale=\singleplotscale]{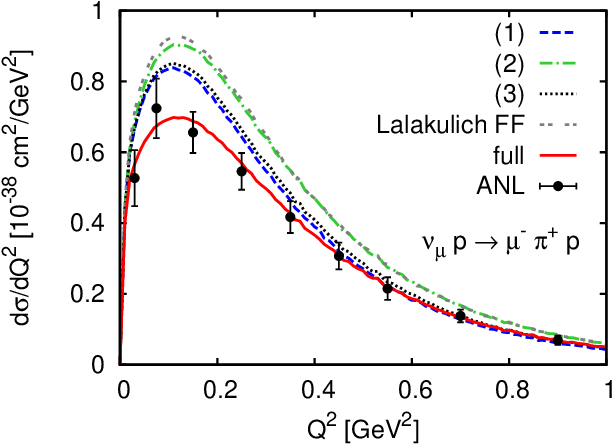}}
  \caption{(Color online) Differential cross section d$\sigma/$d$Q^2$ 
 averaged over the ANL flux taken from
    \cite{Barish:1977qk} as function of $Q^2$ compared to the ANL data taken from Radecky
    \refetal{Radecky:1981fn}. The curves were obtained with different sets of form factors as detailed in the text. Our best result is denoted by the solid line. To compare with data, we applied an invariant mass cut at $W<1.4\GeV$.
    \label{fig:P33_1232_axFF_fit_to_Qs}}
\end{figure}

There have been many earlier attempts to fit these form factors, always under the reasonable assumption that the
vector form factors are precisely known from electron scattering. Most of them adopt the Adler model~\cite{Adler:1968tw} where
\begin{equation}
  C_4^{A}(Q^2)=-\frac{C_5^{A}(Q^2)}{4} \quad \text{and} \quad  C_3^{A}(Q^2)= 0 \eKomma
\end{equation}
thus, one remains with the still unknown $Q^2$-dependence of $C_5^A$ which has to be extracted from data.
A parametrization widely used in the analysis of the neutrino experiments \cite{Kitagaki:1986ct,Kitagaki:1990vs,Barish:1978pj,Radecky:1981fn}
is
\begin{equation}
  C_5^A(Q^2) = {{C_5^A(0)\left[ 1+\frac{a Q^2}{b+Q^2} \right] } {\left( 1+ \frac{Q^2}{{M_A^\Delta}^2}\right)^{-2}}} \eKomma \label{eq:axialformfactorform}
\end{equation}
with $a=-1.21$ and $b=2\GeVsq$ \cite{Adler:1968tw} with an axial mass fitted to the ANL data of $M_A^\Delta=0.98\GeV$ \cite{Radecky:1981fn}.
The cross section d$\sigma/$d$Q^2$ obtained with this form factor parametrization is shown by the dashed line \reffig{fig:P33_1232_axFF_fit_to_Qs} (labeled with (1)) in comparison to the ANL data. It clearly overestimates the data even though the axial form factor was originally fitted to exactly those.

Different parametrizations for the axial form factor have been used, e.g.~in \refcite{Paschos:2003qr,Lalakulich:2006sw}, who use also a modified dipole
\begin{equation}
C_5^{A}(Q^2)=C_5^A(0) \left(1+\frac{Q^2}{3 M_A^2}\right)^{-1} \left( 1+\frac{Q^2}{M_A^2} \right)^{-2} \eKomma
\label{eq:PaschosAx}
\end{equation}
and take $M_A=1.05\GeV$. Using this in combination with our MAID based vector form factors one obtains the dash-dotted line in \reffig{fig:P33_1232_axFF_fit_to_Qs} labeled with (2). Even reducing the axial mass to $M_A=0.95\GeV$ it is not possible to get
a good description of the ANL data (cf.~dotted curve labeled with (3)) without reducing $C_5^A(0)$.
Finally, the double-dashed curve denotes the calculation performed with the form factor set
both vector and axial, of Lalakulich~\refetal{Lalakulich:2006sw}\footnote{Here, the agreement at $Q^2=0$ is better than in the spin 1/2 case; cf.~the discussion of the differences at the end of \refsec{sec:excspinhalf}.}, which also uses the parametrization of \refeq{eq:PaschosAx} for $C_5^A(0)$.
Also in this case, no satisfactory agreement with the data can be reached. Note that in this work, the improvement on the vector form factors was not applied to refit the axial form factors.

\begin{figure}
  \centerline{\includegraphics[scale=\singleplotscale]{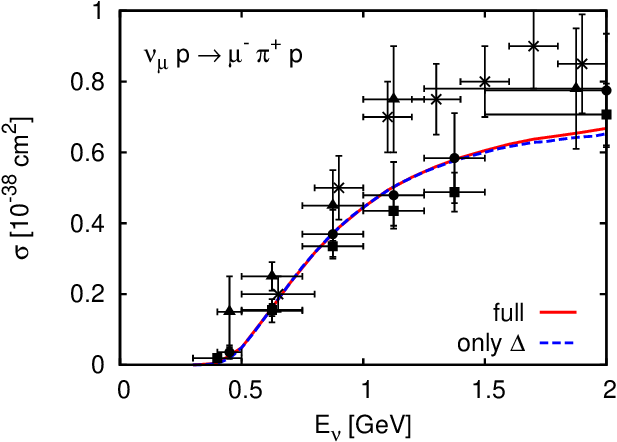}}
  \caption{(Color online) Total $\nu_\mu p \to \mu^- \pi^+ p$ cross section as a function of the neutrino energy compared to the pion
    production data of of ANL (Refs.~\cite{Barish:1978pj} ($\bullet$),
    \cite{Radecky:1981fn} ($\blacksquare$) and \cite{Campbell:1973wg} ($\blacktriangle$))
    and BNL (\cite{Kitagaki:1986ct} ($\times$)). One can see that the above shown isospin 3/2 channel
    is totally dominated by the $\Delta$ resonance (dashed vs.~solid line). No cut on the invariant mass is applied.
    \label{fig:Delta_sigmatot}}
\end{figure}

Comparing these scenarios to the ANL data, we conclude, that improving on the vector form factors without readjusting the axial ones results in a worse description of the data. A refit of the axial form factor is therefore necessary: We rely on the Adler model, and use \refeq{eq:axialformfactorform} as a starting point for our fit of the $Q^2$-dependence of $C_5^A$. We further assume that PCAC holds, which means that we do not take $C_5^{A}(0)$ as a free parameter. As we rely
on PCAC for all other resonance couplings (where no data are available) we prefer to keep it also here.
In addition, this coupling was extracted from the BNL data in \refcite{Alvarez-Ruso:1998hi} and found to be consistent with the PCAC prediction; also recent lattice results support the off-diagonal Goldberger-Treiman relation \cite{Alexandrou:2006mc}.
We find $a=-0.25$, $b=0.04\GeVsq$ and $M_A^\Delta=0.95\GeV$ as best values which is used in the following  as our standard parameter set. Comparing the solid line in \reffig{fig:P33_1232_axFF_fit_to_Qs} to the data, one finds good agreement for our new parameters. The result for the total cross section is
shown in \reffig{fig:Delta_sigmatot} where we compare to the available data. Also there
the agreement with the ANL data is satisfactory.

We stress that we have neglected a non-resonant background which is small in the isospin 3/2 channel $\nu p \to \mu^- \pi^+ p$. Both, Sato~\refetal{Sato:2003rq} and Hernandez
\refetal{Hernandez:2007qq} find within their microscopic models for the non-resonant pion background a correction of the order of $10\%$ in this channel while the discrepancy caused by the difference between the old and the new axial form factor sets is of the order of $30\%$.


\textbf{\res{D}{13}{1520}, \res{D}{33}{1700} and \res{P}{13}{1720}.}
As the $Q^2$-dependence of the axial form factor can be extracted from data only for the \res{P}{33}{1232}, we assume for
the other resonances a simple dipole behavior as done for the $I=1/2$ resonances,
\begin{equation} {C_5^A}(Q^2)=C_5^A(0) \left(1+\frac{Q^2}{{M_A^*}^2}\right)^{-2} \eSemikolon
  \label{eq:axialspinthreehalfff}
\end{equation}
with $M_A^*=1\GeV$. The couplings $C_5^A(0)$ are summarized in \reftab{tab:included_resonances}.  $C_3^A$ and $C_4^A$ are expected to be minor and thus are neglected.

Finally, in \reftab{tab:compspinthreehalf_vecff} we compare our form factors for the \res{D}{13}{1520} resonance to the recent compilation of Lalakulich \refetal{Lalakulich:2006sw}. The agreement is better than in the spin 1/2 case which we compared in \reftab{tab:compspinhalf_vecff}: Focussing on the dominating form factor $C_3$, we still do not agree in the sign of the EM form factors for reasons already discussed at the end of the previous section. However, in this case no relative sign difference remains for the CC form factors as it was in the case of the \res{P}{11}{1440}.
\begin{table}[t]
\begin{center}
\vspace{1cm}
\begin{tabular}{ c c c c}
\noalign{\vspace{-8pt}}
\hline \hline
 &  &\hspace{5pt} this work \hspace{5pt} & Lalakulich \\
\hline
\res{D}{13}{1520} & $C^p_3(0)$ & $-2.70 $ &  $ \phantom{-}2.95 $  \\
                  & $C^n_3(0)$ & $\phantom{-}0.28$  &  $ -1.13 $   \\
                  & $C^V_3(0)$ & $-2.98$ &  $ -4.08 $   \\
                  & $C^p_4(0)$ & $\phantom{-}2.62 $ &  $ -1.05 $  \\
                  & $C^n_4(0)$ & $-1.59$ &  $ \phantom{-}0.46 $   \\
                  & $C^V_4(0)$ & $\phantom{-}4.21$  &  $ \phantom{-}1.51$   \\
                  & $C^p_5(0)$ & $-1.17$ &  $ -0.48 $  \\
                  & $C^n_5(0)$ & $\phantom{-}1.96$  &  $ -0.17 $   \\
                  & $C^V_5(0)$ & $-3.13$ &  $ \phantom{-}0.31$   \\
                  & $C^A_5(0)$ & $-2.15$ &  $ -2.10$ \\
\hline \hline
\end{tabular}
\end{center}
\caption{Comparison of our form factors for the \res{D}{13}{1520} to the ones compiled by Lalakulich \refetal{Lalakulich:2006sw} taken at $Q^2=0$. \label{tab:compspinthreehalf_vecff}}
\end{table}

\subsubsection{\texorpdfstring{Resonances with spin $>$  3/2}{Resonances with spin > 3/2}}

Any formalism describing resonances with spin greater than 3/2 is highly complicated \cite{Shklyar:2004ba}. We
thus make as simplified assumption that all resonances with spin $>$ 3/2 can be treated with the spin 3/2 formalism. As we will show their contributions are anyway negligible in the energy region of interest in this work.

\subsubsection{Results}

\begin{figure*}
  \centerline{\includegraphics[scale=\doubleplotscale]{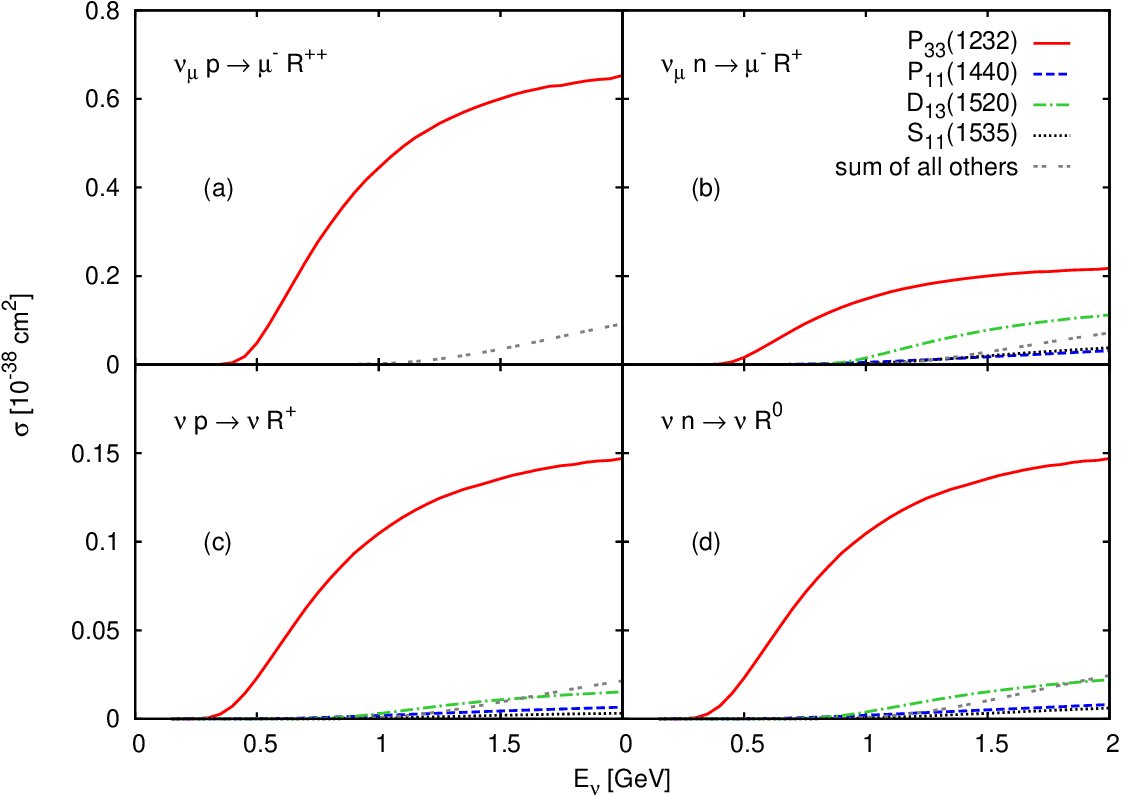}}
  \caption{(Color online) Integrated cross section for CC (panels (a) and (b)) and NC (panels (c) and (d)) induced resonance production on the proton (panels (a) and (c)) and on the neutron (panels (b) and (d)). The different lines indicate the different resonances. Shown are only the results for the four lowest lying resonances --- the sum of the remaining 9 resonances gives the double-dashed line. In the case of CC scattering on protons (panel (a)) only the excitation of isospin 3/2 resonances is possible.
    \label{fig:res_sigmatot}}
\end{figure*}

In \reffig{fig:res_sigmatot}, we show the integrated cross section for CC (panels (a) and (b)) and NC (panels (c) and (d)) induced resonance production on the proton (panels (a) and (c)) and on the neutron (panels (b) and (d)). The dominant contribution comes from the excitation of the $\Delta$ resonance (solid line).


\subsection{\texorpdfstring{Single-$\pi$ non-resonant background}{Single-pion non-resonant background}}
\label{ch:singlepionbackground}

In the preceding section, we have described how we obtain the needed resonance contributions. Since it is well known that photo-nuclear data also contain background contributions we now specify their properties. It is essential to realize that the background contribution must be known for all kinematic variables if one is interested in calculating also differential cross sections.

In the case of neutrinos, the need of such a contribution is justified by the fact that we underestimate the total pion production cross section in the isospin 1/2 channels. We know from the comparison to ANL and BNL data that the resonance excitation alone does not account for all the pion strength (cf.~the discussion in our earlier work~\cite{Leitner:2006ww}). However, the discrepancy is small compared to the total pion production cross section which is dominated by the isospin 3/2 channel.

Let us consider the single-$\pi$ non-resonant background cross section $\dd \sigma_\subBG$. It includes
vector, axial and also interference contributions
\begin{align}
  \dd \sigma_\subBG & = \dd \sigma_\subBG^{\text{V}} + \dd \sigma_\subBG^{\text{A}}+ \dd \sigma_\subBG^{\text{V/A}} \\
  & = \dd \sigma_\subBG^{\text{V}} + \dd \sigma_\subBG^{\text{non-V}} \ePunkt
\end{align}
The vector part is fully determined by electron scattering data. The axial and the
interference term collected under the label ``non-V'' are only present in neutrino scattering and will be fitted to the available neutrino data.

\subsubsection{Vector part}

We first discuss the vector part of the non-resonant pion cross section, $\dd \sigma_\subBG^{\text{V}}$.
Our strategy will be to evaluate the unpolarized pion production cross section $\ell(k) N(p) \to \ell'(\kpr) \pi(\kpi) N(\ppr)$, $\dd\sigma_{N \pi}^{\text{V}}$, and to subtract afterwards the resonance contribution.
We understand this subtracted contribution, which then includes also the resonance-background
interference terms, as a single-$\pi$ background denoted as $\dd\sigma_\subBG^{\text{V}}$.
We are actually assuming that the resonances do not interfere among themselves or with the background, so that
\begin{multline}
  \frac{\dd\sigma_\subBG^{\text{V}}}{\dd \omega \dd\Omega_{\kpr} \dd\Omega_{\kpi} }  =  \\
   \frac{\dd\sigma_{N \pi}^{\text{V}}}{\dd \omega \dd\Omega_{\kpr} \dd\Omega_{\kpi} }
  -\sum_{R}\frac{\dd\sigma_{\ell N\to \ell R\to \ell N\pi}^{\text{V}}}{\dd \omega \dd\Omega_{\kpr} \dd\Omega_{\kpi} } \ePunkt
  \label{eq:piBG}
\end{multline}

The first term of the rhs, the cross section for $\ell(k) N(p) \to \ell'(\kpr) \pi(\kpi) N(\ppr)$, $\dd\sigma_{N \pi}^{\text{V}}$, is given by
\begin{align}
  \frac{\dd\sigma_{N \pi}^\text{V}}{\dd \omega \dd\Omega_{\kpr} \dd\Omega_{\kpi} } =& \int \frac{\abskpr \abskpi}{512 \pi^5} \; \left[ (k \cdot p)^2 -
    \ml^2 M^2 \right]^{-1/2} \nonumber \\ & \times \delta (\ppr^2 - \Mpr^2) \; | \bar{\ME}_{N \pi} |^2 \dd \kpiz        \label{eq:onepioncross}
\end{align}
with $\ppr=k+p-\kpr-\kpi$, $\abskpr=\sqrt{{\kpr_0}^2-\mlpr^2}$ and
$\abskpi=\sqrt{{\kpiz}^2-\mpi^2}$ and with $\Omega_{\kpi}$ denoting the solid angle of
the pion.  The $\delta$-function $\delta (\ppr^2 - \Mpr^2)$ eliminates the $\dd \kpiz$ integration.
The hadronic tensor $H^{\mu \nu}_{N \pi}$ entering the matrix element in \refeq{eq:onepioncross} is written in a form similar to that of resonance production
\begin{equation}
  H^{\mu \nu}_{N \pi} = \frac12 \Tr \left[ (\slashp + M ) \tilde{\mathcal{V}}^{\mu}_{N \pi} (\slashp' + M' ) \mathcal{V}^{\nu}_{N \pi} \right] \eKomma
\end{equation}
with
\begin{equation}
  \tilde{\mathcal{V}}^{\mu}_{N \pi}=\gamma_0 {\mathcal{V}^{\mu}_{N \pi}}^{\!\!\!\!\dagger} \gamma_0 \ePunkt
\end{equation}

The hadronic vertex can be parametrized in the most general way~\cite{Berends:1967vi} as
\begin{equation}
  {\mathcal{V}_{N \pi}^\mu}=\sum_{i=1}^{6} A_i^{N \pi} M_i^\mu \eKomma \label{eq:EMpionbackgroundcurrent}
\end{equation}
with (in the notation of MAID \cite{Pasquini:2007fw})
\begin{align}
M_1^\mu &=\frac{-\ii}{2}\gamma^5 (\gamma^{\mu}\slashed{q} -\slashed{q} \gamma^{\mu})=-\ii\gamma^5 (\gamma^{\mu}\slashed{q} -q^\mu)   \; ,\nonumber \\
M_2^\mu &=2 \ii  \gamma^5 \left(P^\mu q\cdot\left(k_\pi-\frac{q}{2}\right)-P \cdot q \left(k_\pi-\frac{q}{2}\right)^\mu\right)  \; , \nonumber \\
M_3^\mu &=  -\ii \, \gamma^5 (\gamma^{\mu} k_\pi\cdot q-\slashed{q}k_\pi^{\mu})  \; ,\label{eq:pionbackground_Mi} \\
M_4^\mu &=-2\ii \, \gamma^5 (\gamma^{\mu} q\cdot P-\slashed{q} P^\mu) -2 M_N M_1^\mu  \; ,\nonumber \\
M_5^\mu &=  \ii  \, \gamma^5 (q^\mu k_\pi\cdot q - q^2 k_\pi^\mu)    \; ,\nonumber \\
M_6^\mu &=-\ii   \, \gamma^5 (\slashed{q} q^{\mu}-q^2 \gamma^{\mu})  \nonumber
\end{align}
and $P^\mu=(p+\ppr)^\mu/2$.  The so-called invariant amplitudes $A_1^{N \pi}, \ldots, A_6^{N \pi}$, depending both on the probe and the reaction channel, are functions of three scalars which completely determine all the $4$-vectors at the vertex. We choose
$W=\sqrt{s}$, $Q^2=-q_\mu q^\mu$ and the CM scattering angle $\theta$ between $\myvec{q}$
and $\myvec{\kpi}$ as such a set of independent scalars.

The second part of \refeq{eq:piBG} is obtained in the following way: the resonances are assumed to decay isotropically in their rest-frame, i.e.
\[
\frac{\dd\Gamma_{R\to N\pi}}{\dd\Omega_{\kpi}^\text{CM}}=\frac{\Gamma_{R\to N\pi}}{4\pi}             \eKomma
\]
and consequently the single resonance contributions are given by
\begin{multline}
 \frac{\dd\sigma_{\ell N\to \ell R\to \ell N\pi}^{\text{V}}}{\dd \omega \dd\Omega_{\kpr} \dd\Omega_{\kpi} } =
      \frac{\dd\sigma_R^\text{V}}{\dd \omega \dd\Omega_{\kpr}} \frac{1}{4 \pi} \frac{\Gamma_{R\to N\pi}}{\Gamma_{R}}
    \frac{d\Omega_{\kpi}^\text{CM}}{d\Omega_{\kpi}} \; .
\end{multline}
The vector part of the
resonance cross section has been introduced in the previous section and the solid-angle transformation is given by~\cite{Byckling}
\begin{multline}
\frac{d\Omega_{\kpi}^\text{CM}}{d\Omega_{\kpi}}=    \frac{\sqrt{\ppr^2}\myvec{\kpi^\text{2}}}{|\myvec{\kpi^{\tinyMM{CM}}}|\left(|\myvec{\kpi}|\ppr^0-|\myvec{\ppr}|
      \kpi^0 \cos(\theta_\pi)\right)}
\end{multline}
where $\theta_\pi=\measuredangle(\myvec{\kpi},\myvec{\ppr})$.

\textbf{Electron scattering.} In the case of electron scattering, the possible channels are
\begin{align}
  e^- p &\to e^- p \pi^0 \eKomma\\
  e^- p &\to e^- n \pi^+ \eKomma\\
  e^- n &\to e^- n \pi^0 \eKomma\\
  e^- n &\to e^- p \pi^- \ePunkt  \label{eq:EMpionbackgroundchannels}
\end{align}
For each of these channels, a set of invariant amplitudes $A_i^{N \pi, \subEM}$ can be fitted to data. We use the MAID parametrization~\cite{MAIDWebsite,Tiator:2006dq,Drechsel:1992pn}.

\textbf{Charged current scattering.} In the case of charged current neutrino scattering,
there are three pion production channels, namely
\begin{align}
  \nu p &\to \lep^- p \pi^+ \eKomma\\
  \nu n &\to \lep^- n \pi^+ \eKomma\\
  \nu n &\to \lep^- p \pi^0 \ePunkt
\end{align}
Applying isospin relations to relate $A_i^{N \pi, \subCC}$ to the known $A_i^{N \pi,
  \subEM}$, one gets (for details, see Appendix \ref{sec:isoBG})
\begin{align}
  A_i^{p \pi^+, \subCC} &= \sqrt{2} A_i^{n \pi^0, \subEM} +  A_i^{p \pi^-, \subEM}  \eKomma \\
  A_i^{n \pi^+, \subCC} &= \sqrt{2} A_i^{p \pi^0, \subEM} -  A_i^{p \pi^-, \subEM}  \eKomma \\
  A_i^{p \pi^0, \subCC} &= A_i^{p \pi^0, \subEM} -A_i^{n \pi^0, \subEM}-\sqrt{2} A_i^{p \pi^-, \subEM} \eKomma
\end{align}
so that the CC vector part is fully determined.

\textbf{Neutral current scattering.}  The following channels contribute to neutral current pion production
\begin{align}
  \nu p &\to \nu p \pi^0 \eKomma\\
  \nu p &\to \nu n \pi^+ \eKomma\\
  \nu n &\to \nu n \pi^0 \eKomma\\
  \nu n &\to \nu p \pi^- \ePunkt
\end{align}
Isospin relations required to relate the invariant amplitudes $A_i^{N \pi, \subNC}$ to the known $A_i^{N \pi, \subEM}$ are more complicated than in the CC case and can be found, e.g., in Section III of \refcite{Hernandez:2007qq}.

\subsubsection{Non-vector part}

In the case of neutrino scattering, an \emph{a priori} unknown non-vector background is present.
In principle, one can write down a similar expression as \refeq{eq:EMpionbackgroundcurrent} for the
axial current. This would lead to a large number of unknown amplitudes $A_i$ to be fixed with data. However, the scarcity of available data makes this option impracticable.

Recently, microscopic models for the elementary reaction have been developed \cite{Sato:2003rq,Hernandez:2007qq} including besides non-resonant terms only the $\Delta$ as intermediate resonance state. However, as pointed out before, the background contribution to the \emph{total} pion production cross section is small because of the dominant isospin 3/2 channel. Thus, in the present work, we rather use a simple ansatz for the non-vector background expecting that the final results in nuclei are not sensitive to the background details.

\begin{figure}
  \centerline{\includegraphics[width=0.45\textwidth]{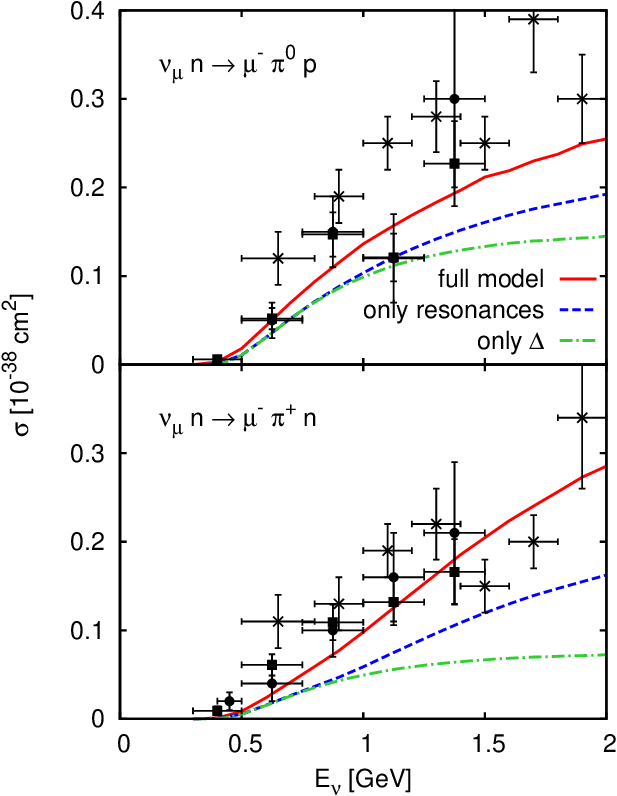}}
  \caption{(Color online) Total CC pion production cross sections for the mixed isospin channels as a
    function of the neutrino energy compared to the pion production data of of ANL
    (Refs.~\cite{Barish:1978pj} ($\bullet$) and \cite{Radecky:1981fn} ($\blacksquare$)) and BNL (\cite{Kitagaki:1986ct}
    ($\times$)). The solid lines denote the our full result including the non-resonant
    background following \refeq{eq:non-resBG} with $b^{p \pi^0}=3$ and $b^{n \pi^+}=1.5$.
    Furthermore, we show the results for pion production only through the excitation and the subsequent decay of all resonances (dashed lines) or through the $\Delta$ alone (dash-dotted lines). No cut on the invariant mass is applied. \label{fig:xsec_pionprod_wBG}}
\end{figure}

\textbf{Charged current scattering.}
As already discussed in \refsec{sec:excitation_spin_threehalf}, we neglect a non-resonant background in the isospin 3/2 channel $\nu p \to \mu^- \pi^+ p$ where it is only a small correction. Furthermore, we assume that $\dd \sigma_\subBG^\text{V}$ and $\dd
\sigma_\subBG^\text{non-V}$ have the same functional form, i.e.,
\begin{align}
  \dd {\sigma_\subBG} = \dd \sigma_\subBG^\text{V} + \dd \sigma_\subBG^\text{non-V} = (1+b^{N \pi})\; \dd \sigma_\subBG^\text{V} \eKomma
  \label{eq:non-resBG}
\end{align}
where the global factor $b^{N \pi}$ depends on the process, $\nu n \to \lep^- n \pi^+$ or
$\nu n \to \lep^- p \pi^0$. The data sets for the two CC scattering channels off neutrons allow then the fit of the two parameters.
With $b^{p \pi^0}=3$ and $b^{n \pi^+}=1.5$ a reasonable
agreement with the ANL data is achieved as can be seen from \reffig{fig:xsec_pionprod_wBG} where the solid line denotes our full calculation. This figure emphasizes again the need for a non-resonant background in the isospin 1/2 channel. Pion production only through the $\Delta$ (dash-dotted lines) is not sufficient to describe the data. The inclusion of higher resonances increases the pion production cross section (dashed lines), but still, the non-resonant background is required (solid lines).

Our numbers are in agreement with general isospin considerations. The dominant
contribution to the non-resonant background comes from the nucleon-pole term \cite{valverdeprivcomm,Hernandez:2007qq}, an isospin
1/2 channel. Using Clebsch-Gordon coefficients, we find for the yields $\pi^0/\pi^+ = 2$,
which nicely corresponds to $b^{p \pi^0}/b^{n \pi^+}=2$.

\begin{figure}
  \centerline{\includegraphics[width=0.45\textwidth]{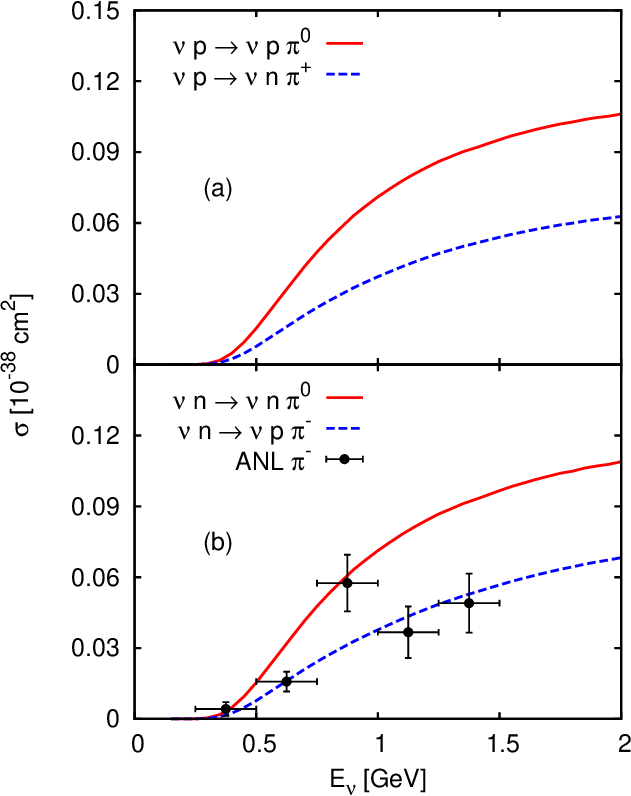}}
  \caption{(Color online) Total NC cross sections for pion production through the
    excitation and the subsequent decay of all included resonances as a function of the
    neutrino energy. Panel (a) shows the cross sections on protons, (b) the ones on
    neutrons.  In addition, we show the data of ANL \cite{Derrick:1980nr} which have to be
    compared with the $\nu n \to \nu p \pi^-$ channel (dashed line in panel (b)). No cut
    on the invariant mass is applied. \label{fig:xsec_pionprod_NC}}
\end{figure}

\textbf{Neutral current scattering.}
Pion data for NC scattering are even more scarce
than for the CC case. The available data is compared in \reffig{fig:xsec_pionprod_NC} to our results for the resonance induced pion production. We find a good agreement --- dashed line in panel (b) compared to data --- already without non-resonant background. In view of this, we abstain from fitting the non-vector part to this data and neglect a NC background.

\section{Lepton-nucleus interactions within the GiBUU model}

Over the last two decades, the Giessen theory group has developed a  Boltzmann-Uehling-Uhlenbeck (BUU) transport model for nuclear reactions, named Giessen BUU (GiBUU) \cite{gibuu,gibuupaper}.
Besides its original field of application, which were heavy-ion collisions~\cite{Larionov:2003av,Larionov:2005kz,Wagner:2004ee}, the model was also successfully applied to high-energy non-resonant electron-induced reactions~\cite{Falter:2004uc,Gallmeister:2005ad}, photon-, pion- and electron-induced processes in the resonance region \cite{Lehr:1999zr,Lehr:2003km,Muhlich:2003tj,Muhlich:2006ps,buss_phd}, and neutrino-induced reactions \cite{Leitner:2006ww,Leitner:2006sp}. In April 2008, the GiBUU model source code has been published under GNU General Public License in a restructured modern form for public use~\cite{gibuu}.

The GiBUU transport approach \cite{buss_phd,gibuu} gives a microscopic description of the final state process. Thereby all kind of coupled channel rescattering effects (e.g.~charge exchange processes, resonance production and decays in the medium) are included. All particles are propagated in hadronic mean fields and full offshell propagation of the hadronic resonances has been implemented.

In the following, we consider the reactions of electrons and neutrinos with nuclei in the impulse approximation. Thus, we assume that the elementary projectile scatters off one single nucleon bound in the nucleus. In the \emph{initial step} the projectile excites a resonance, generates a quasielastic event or a non-resonant background process. For exclusive or semi-inclusive reactions the resulting particle yield of this initial state must then be propagated through the nucleus. Such \emph{final state interactions} can then be treated within the transport approach. However, in this work we consider only inclusive cross sections and postpone the treatment of semi-inclusive reactions to a forthcoming
publication~\cite{InProgress}.

\subsection{Nuclear ground state}
\label{sec:groundstate}
The phase space density of the nucleons bound in a nucleus is treated within a local Thomas-Fermi (LTF) approximation, i.e., at each space point the nucleon momentum distribution is given by a Fermi sphere
\begin{equation}
  \label{oneParticleFSD}
  f^{n,p}(\myvec{r},\myvec{p})= \Theta\left(p_F^{n,p}(\myvec{r})-\left|\myvec{p}\right|\right) \ePunkt
\end{equation}
 with radius $p^{n,p}_F=(3\pi^2\rho^{n,p})^{1/3}$. The normalization is chosen such that the particle density is retrieved by
\begin{equation}
  \rho^{n,p}(\myvec{r})=g~  \int f^{n,p}(\myvec{r},\myvec{p})~ \frac{\dd^3\myvec{p}}{(2\pi)^3} \eKomma
\end{equation}
where  $g=2$ denotes the spin degeneracy factor. In the LTF approximation, the Pauli blocking factor is given by
\begin{equation}
  P_{\mbox{\tiny{PB}}}^{n,p}(\myvec{r},\myvec{p})=1-f^{n,p}(\myvec{r},\myvec{p}) \; . \label{Pauli}
\end{equation}

\subsubsection{Density profiles}
The density profiles $\rho(r)$ are implemented according to a parametrization collected in \refcite{osetPionicAtoms}: the proton density is based
on the compilation of \refcite{DeJager:1974dg} from electron scattering; the neutron
density is provided by Hartree-Fock calculations.

\subsubsection{Mean field potentials}
Nucleons (and also resonances and mesons) in the medium are exposed to hadronic mean-field potentials.

The nucleon mean-field potential is defined in the nucleus rest frame and parametrized according to Welke \refetal{Welke} as a sum of a Skyrme term depending only on density and a momentum-dependent contribution of Yukawa-type interaction
\begin{align}
  \label{nucPot}
  V_{\mathrm{N}}(\myvec{p},\myvec{r}) = & A\ \frac{\rho(\myvec{r})}{\rho_{0}}+
  B\ \left(\frac{\rho(\myvec{r})}{\rho_{0}}\right)^{\tau} \nonumber \\
  & +\frac{2 C}{\rho_{0}} \int \frac{\dd^{3}\myvec{p}'}{\left(2\pi\right)^{3}} \frac{g
    \left(f^n(\myvec{r},\myvec{p}^{\, \prime})+f^p(\myvec{r},\myvec{p}^{\, \prime})\right)}
  {1+\left(\frac{\myvec{p}-\myvec{p}^{\, \prime}}{\Lambda}\right)^{2}} \ePunkt
\end{align}
Different possible parameter sets for the latter potential have been fixed by Teis \refetal{Teis:1996kx} analyzing nucleon-nucleus scattering. The ones being used in this work are given in \reftab{potParTable}: parameter set \#1 gives a \emph{momentum dependent} potential with a nuclear matter compressibility $C=290 \MeV$, while set \#2 results in a \emph{momentum independent} one with $C=215 \MeV$.
\begin{table}
\begin{center}
\vspace{1cm}
\begin{tabular}{c c c c c c}
  \noalign{\vspace{-8pt}}
  \hline \hline
  Set  & $A$ [MeV] & $B$ [MeV] & $C$ [MeV] & $\tau$ &  $\Lambda$ [fm$^{-1}$]    \\
\hline
  \#1 &  $-29.3$  &  $57.2$ &  $-63.5$ &$1.76$ & $2.13$   \\
  \#2 &  $-287.$  &  $234.$ &  $0$     &$1.23$ & --    \\
\hline \hline
\end{tabular}
\end{center}
\caption{Parameter sets for the Skyrme parametrization of \refeq{nucPot} (for explicit details see \refcite{Teis:1996kx}). Set \#1 is momentum dependent whereas set \#2 is momentum independent.}
\label{potParTable}
\end{table}
Photon-nucleus interactions indicate that the $\Delta$(\res{P}{33}{1232})-resonance potential has a depth of about $-30\MeV$ at $\rho_{0}$\,\cite{Ericson:1988gk,Peters:1998mb}. Comparing to a
momentum independent nucleon potential, which is approximately $ -50\MeV$ deep, the
$\Delta$ potential is, therefore, approximated by
\begin{equation}
  V_{\Delta}(\myvec{p},\myvec{r})=\frac{2}{3}\ V_{\mathrm{N}}(\myvec{p},\myvec{r})\ePunkt
  \label{DeltaPotential}
\end{equation}
We assume for all spin 3/2 resonances the same potential as for the $\Delta$-resonance.
For all spin 1/2 and spin $> 3/2$ resonances  we assume the same potential as for the
nucleon. Furthermore, we found that the impact of a mean field potential acting on the pion~(for details cf.~\refcite{Buss:2006vh}) is negligible for the later regarded processes.

For convenience, one can rewrite the mean field potentials as scalar potentials $U_s$ which are defined by
\begin{align}
\sqrt{\myvec{p}^2+M_0^2}+V(\myvec{p},\myvec{r})=\sqrt{\myvec{p}^2+(M_0+U_s(\myvec{p},\myvec{r}))^2}
  \label{scalarPot}
\end{align}
with the Breit-Wigner vacuum mass $M_0$. In this way one introduces effective masses for the incoming nucleon $M=M_N+U_s(\myvec{p},\myvec{r})$ and for the outgoing baryons $\Mpr=M_B+{U_s}_B(\myvec{\ppr},\myvec{r})$ where $\myvec{p}$ and $\myvec{\ppr}$ are the corresponding momenta.

\subsubsection{Momentum densities}
\label{sec:momDensities}
\begin{figure}
   \centerline{\includegraphics[scale=\singleplotscale]{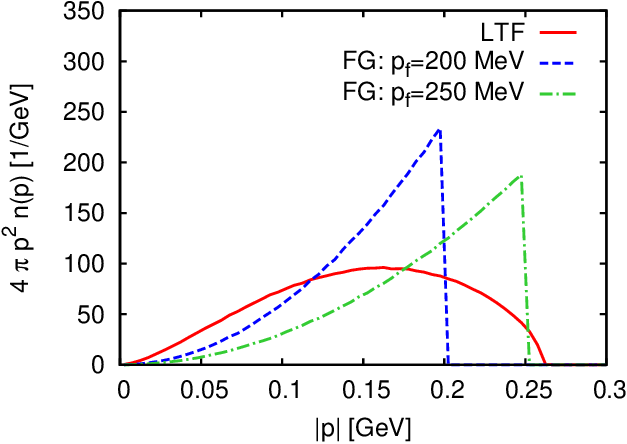}}
  \caption{(Color online) Momentum density distribution of protons in $^{16}$O. The solid curve
    represents the LTF result with the density parametrization of
    \refcite{osetPionicAtoms}. The dashed and  dotted curves show the results for
    a global Fermi momentum of 0.2 GeV and 0.25 GeV according to \refeq{constMomDens}. The normalization condition is given by $4\pi\int \dd |\myvec{p}| |\myvec{p}|^2 n(|\myvec{p}|) =A$.
  }
  \label{momDist}
\end{figure}

\begin{figure}
   \centerline{\includegraphics[scale=0.75]{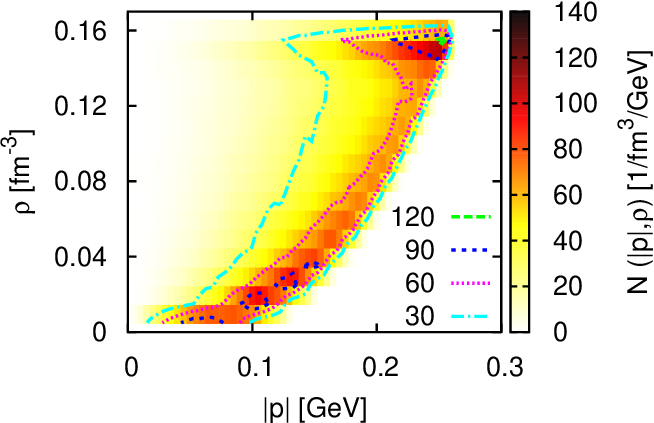}}
  \caption{(Color online) Probability density $N(|\myvec{p}|,\rho)$ of finding a nucleon with absolute momentum $|\myvec{p}|$ at a position with density $\rho$ in an {\oxygen} nucleus within the LTF scheme.  The normalization condition is $\int \dd\rho\, \dd |\myvec{p}| N(|\myvec{p}|,\rho)=1$.}
    \label{momDensDist}
\end{figure}

With a given density parametrization, the single-particle phase-space density $f(\myvec{r},\myvec{p})$ is fully determined. The momentum density is then given by
\begin{equation}
  n^{n,p}(\myvec{p})=g\int f^{n,p}(\myvec{r},\myvec{p}) \frac{\dd^3\myvec{r}}{(2\pi)^3} \ePunkt
\end{equation}
\reffig{momDist} shows our result for $n^{p}(\myvec{p})$ in {\oxygen} as a function of the absolute momentum $p$. To point out the differences of the LTF approximation and the common \emph{global Fermi gas (GFG)} approximation the figure shows also the GFG results for comparison. In the GFG approximation one assumes that the nucleus is just a sphere with radius $R$, the one particle phase space densities are given by
\begin{equation}
  f^{n,p}_{\mbox{\tiny{GFG}}}(\myvec{r},\myvec{p})=\Theta(R-r)\Theta(p^{n,p}_f-p)
\label{eq:phaseSpace}
\end{equation}
and the momentum densities are simple step-functions
\begin{align}
  n^n_{\mbox{\tiny{GFG}}}(p)&=\frac{A-Z}{\frac{4}{3}\pi (p^n_f)^3} \Theta(p^n_f-p)  \nonumber\\
  n^p_{\mbox{\tiny{GFG}}}(p)&=\frac{Z}{\frac{4}{3}\pi (p^p_f)^3} \Theta(p^p_f-p) \; .
  \label{constMomDens}
\end{align}
In \reffig{momDist}, it is shown that the realistic density parametrization of \refcite{osetPionicAtoms} leads within the LTF ansatz to a less peaked momentum distribution and to more strength at low momenta compared to the global Fermi gas approximation. Additionally, the LTF approximation generates  a  space-momentum correlation for the phase space density, which does not exist for the GFG approximation: in \reffig{momDensDist} one observes that the probability $N(p,\rho)$ to find a nucleon  with absolute momentum $p$ in a nuclear environment of density $\rho$ has a ridge structure. The low-momentum nucleons experience a low density environment while the high momentum nucleons tend to sit in a high density environment. This feature will become important in the later discussion of inclusive electron scattering.

\subsubsection{Spectral functions and in-medium widths}

The spectral function of an outgoing particle with four-momentum $p$ is given by
\begin{equation}
  \mathcal{A}(p)=\frac{-\ImaginaryPart \Sigma(p)}{(p^2-M_0^2-\RealPart \Sigma(p))^2+(\ImaginaryPart \Sigma(p))^2}  \label{eq:inmedspecfunc}
\end{equation}
where $\Sigma$ denotes the self energy.

To deduce the imaginary part of the self energy of a particle in the medium, directly related to its width, we have to consider the modification of its free decay width due to the Pauli blocking of final state nucleons and also collisional broadening due to interactions with the surrounding nucleons.

To estimate this collisional broadening of a particle with momentum $p$ and energy $E$, we employ the low-density approximation
\begin{align}
  \Gamma_\text{coll}(p_0,\myvec{p},\myvec{r})
  =&\sum_{i=n,p}\int f^i(\myvec{p}',\myvec{r}) \left. \sigma_i(p_0,\myvec{p},\myvec{p}') \right. \nonumber\\ &\times \left.~v_\text{rel}(p_0,\myvec{p},\myvec{p}') \right.~\dd^3\myvec{p}' \ePunkt
\end{align}
Here the variable $v_\text{rel}$ denotes the relative velocity of the regarded particle and a nucleon with momentum $\myvec{p}'$; the nucleon phase space densities $f^i$ have been introduced in \refeq{eq:phaseSpace}. The total nucleon-particle scattering cross sections $\sigma_i$ are chosen according to the GiBUU collision term~\cite{gibuu}. Altogether, the imaginary part of the self energy is in the rest-frame of the particle given by
\begin{align}
  \ImaginaryPart \Sigma(p_0,\myvec{p},\myvec{r}) &=-\sqrt{p^2} \left \lbrace \Gamma_\text{free,Pauli blocked}\left(p_0,\myvec{p},\myvec{r}\right) \right.\nonumber \\
&\left.+\gamma \Gamma_\text{coll}\left(p_0,\myvec{p},\myvec{r}\right)\right \rbrace  \eKomma
\label{inMed_Imag}
\end{align}
where $\gamma$ denotes the boost factor from nucleus rest frame to particle rest frame. The vacuum decay widths $\Gamma_\text{free}$ are parametrized according to Manley \refetal{manley} (for details cf.~\refcite{buss_phd}).

Since the self energy is an analytic function of $p_0$, we can use dispersion relations to deduce the off-shell behavior of the real parts from the imaginary part.
We apply a once-subtracted dispersion relation with the on-shell (OS) energy, which is defined by
\eb
p_0^\text{OS}=\sqrt{\myvec{p}^2+M_0^2+\RealPart \Sigma(p_0^\text{OS},\myvec{p},\myvec{r})} \eKomma
\ee
as subtraction point. This yields
\begin{align}
\label{eq:disp}
\RealPart \Sigma(p,\myvec{r})&=\RealPart \Sigma(p_0^\text{OS},\myvec{p},\myvec{r})+\frac{p_0-p_0^\text{OS}}{\pi}~ \nonumber\\
&\times\wp \int^\infty_{-\infty} \dd p_0' \frac{\ImaginaryPart \Sigma(p_0',\myvec{p},\myvec{r})}{(p_0'-p_0^\text{pole})(p_0'-p_0)} +\RealPart\, C_\infty  \ePunkt
\end{align}
In the same line as Lehr \refetal{Lehr:2001qy,lehr_phd}, we demand that the in-medium shift of the on-shell energy is determined by the mean-fields
\eb
p_0^\text{OS}=\sqrt{\myvec{p}\;^2+(M_0+U_s(\myvec{p},\myvec{r}))^2} \eKomma
\ee
with the scalar potential $U_s$ defined in \refeq{scalarPot}. Consequently, the non-dispersive contribution to $\RealPart \Sigma$ is given by
\eb
\RealPart \Sigma(p_0^\text{OS},\myvec{p},\myvec{r})= 2M_0 U_s(\myvec{p},\myvec{r})+U_s(\myvec{p},\myvec{r})^2 \ePunkt
\ee
In the numerical realization, we approximate the dispersion integral \refeq{eq:disp} for $\RealPart \Sigma$ by
\begin{align}
\RealPart \Sigma(p,\myvec{r})=&\RealPart \Sigma(p_0^\text{OS},\myvec{p},\myvec{r})+\frac{p_0-p_0^\text{OS}}{\pi}~ \nonumber\\
& \wp \left(\int_{E_\text{min}}^{E_1} \dd p_0' \frac{\ImaginaryPart \Sigma(p_0',\myvec{p},\myvec{r})}{
(p_0'-p_0^\text{OS})(p_0'-p_0)}\right. \nonumber\\
&+\left. \int^{E_2}_{E_1} \dd p_0' \frac{\ImaginaryPart \Sigma(p_0',\myvec{p},\myvec{r}) }{(p_0'-p_0^\text{OS})(p_0'-p_0)}  \frac{E_2-p_0'}{E_2-E_1}\right)
\label{eq:numericsRealPart}
\end{align}
with the cutoff parameters $E_1=5 \GeV$ and $E_2=7 \GeV$. $E_\text{min}$ is determined by the mass of the lightest decay product of the particle. We have checked the dependence of our results on the cutoffs and found only a marginal impact. The whole procedure guarantees analytical self energies and, therefore, normalized spectral functions.

For the spectral function of the initial state nucleon, we consider only the real part of the self energy generated by the mean-field potential and neglect the imaginary part.

\subsection{Lepton-nucleon interaction in the medium}
\label{sec:leptonnucleus}
As outlined in the beginning of this section, the wavelength of the exchanged boson is considered small as compared to intra-nucleon distances in the nucleus such that the nuclear reaction may be treated in impulse approximation. In this picture the lepton interacts with a single nucleon being embedded in the nuclear medium.\footnote{We note that the impulse approximation is expected to work only at high momenta $(|\myvec{p}| \gtrsim 300\MeV)$ \cite{Ankowski:2008df}.}
 Hereby either a nucleon is scattered in a quasielastic event, or a resonance or one direct pion (background contribution) are produced. For these reactions we account for the in-medium-modifications introduced in \refsec{sec:groundstate}.

In the following we discuss how the different contributions to $\dd \sigma_\subtot$ of \refeq{eq:generalDecomposition} get modified when the nucleon is bound inside a nucleus, i.e.,
\begin{equation}
 \dd \sigma_\subtot \to  \dd \sigma_\subtot^{\text{med}},
\end{equation}
where $\dd \sigma_\subtot^{\text{med}}$ includes the nuclear modifications discussed below.

\subsubsection{Quasielastic scattering and resonance excitations}
For the reaction cross sections with a single hadronic final state, e.g., resonance excitation or quasielastic scattering, the implementation of medium-modifications is straightforward. The vacuum spectral function for the resonance in \eqref{eq:RESdoublediff_xsec}
and the one of the outgoing nucleon in \refeq{eq:QEdoublediff_xsec}, namely $\mathcal{A}_\text{vac}^\text{N}(p)=\delta(p^2-M_N^2)$, get replaced by the
in-medium spectral functions
\eb
\mathcal{A}_\text{vac}\to \mathcal{A}_\text{medium}\ePunkt
\ee
The four-momentum $\pr{p}$ of the outgoing hadron is directly given by energy and momentum conservation. $M$ and $\Mpr$ in the corresponding equations turn into effective masses $M \to M + U_S$
due to the presence of the mean field potential.

As detailed by Naus \refetal{Naus:1990em}, the most general in-medium vertex for the quasielastic $\gamma^\star N\to N'$ process has in general 12 linearly independent Lorentz structures, which yields also 12 different form factors. These form factors may, unlike the vacuum ones, depend on all the possible independent Lorentz scalars --- e.g.~$q^2$, $P^2$ and $P\cdot q$ where $P=p+p'$. Unfortunately, there is no feasible way to extract all these different form factors from the available experimental data. Therefore, we model the in-medium nucleon-photon vertex in the same spirit as de Forest \cite{DeForest:1983vc}, which means that the in-medium vertex structure is assumed to be the same as the vacuum one. Furthermore, we assume that the vertex structure and the form factors are not modified by the medium and depend only on $Q^2$.

In the case of quasielastic scattering, the effective masses of in- and outgoing nucleons ($M$ and $\Mpr$) are in general not equal due to the momentum dependence of the mean-field potential. Adding an extra term in the QE current by replacing in \refeq{eq:QEVECcurrent}
\begin{equation}
 \gamma^\mu \to \gamma^\mu + \frac{\slashq q^\mu}{Q^2}
\end{equation}
ensures vector current conservation even when the masses of the initial and final nucleons differ. This extra term vanishes by applying Gordon identities in the limit of free nucleons, so that when $M=\Mpr$, we retrieve the standard expression of \refeq{eq:QEVECcurrent}.

Pollock \refetal{Pollock:1995dz} find that this extra term $\slashed{q} q^\mu / Q^2 $, which guarantees charge conservation at the vertex, can also be generated via a gauge transformation of the result obtained in Landau gauge to Feynman gauge. They argue that the above term is no more than a gauge relict. We have checked the impact of this term on our results finding no noticeable modifications.

Note that in this model the bound state properties of the nucleon are present for both for the initial and final states. We also note that a struck nucleon does not necessarily has to be knocked out of the nucleus, but can also stay bound in the potential. This is different from the model of Benhar \refetal{Benhar:2005dj,Nakamura:2007pj}, where the outgoing nucleon is always assumed to be knocked out
and described in terms of a single parameter, an average removal energy.

As in the quasielastic case, we also assume that for resonance production, the in-medium vertex structure and the form factors are not modified by the medium.

\subsubsection{Single-pion production backgrounds}
Let us now consider the single pion production background channel. The  single pion cross section obtained from MAID is considered as a parametrization of the vacuum data and the background is the difference of data and resonance contributions. Thus we first construct according to \refeq{eq:piBG} the total single pion cross section using the MAID input and the contribution of the resonances to single-pion production using vacuum kinematics (no modifications besides Fermi motion, in particular no potentials, no Pauli blocking). Then we evaluate the difference of both cross sections using vacuum kinematics --- this yields the background cross section $\dd\sigma_\subBG/(\dd \omega \dd\Omega_{\kpr})$.

This background cross section is now assumed not to be influenced by the potentials, which means that the in-medium background cross section is given by
\eb
\frac{\dd\sigma_\text{BG, medium}}
{\dd \omega \dd\Omega_{\kpr} } (p,q)
=\frac{\dd\sigma_\text{BG, vacuum}}
{\dd \omega \dd\Omega_{\kpr} } (p_\text{vac},q)  \eKomma
\ee
where $p_\text{vac}=\left(\sqrt{M^2+\myvec{p}^2},\myvec{p}\right)$.

\section{Inclusive scattering off nuclei}

The inclusive cross section for scattering off a nucleus is given by
\begin{align}
  \dd \sigma_\subtot^{\ell A \to \ell' X} =&g \int_\text{nucleus} \dd^3 r \int \frac{\dd^3 p }{(2\pi)^3} \Theta(p_F(r)-p) \nonumber \\
 &\times  \frac{1}{v_\text{rel}} \frac{k \cdot p}{\kz \pz} \dd \sigma_\subtot^{\text{med}}  P_{\mbox{\tiny{PB}}}(\myvec{r},\myvec{p}) \eKomma
\end{align}
where the kinematical prefactor corrects for the nuclear flux. The Pauli blocking factor $P_{\mbox{\tiny{PB}}}$ is given in \refeq{Pauli}. The relative velocity between lepton and nucleus is $v_\text{rel} \approx 1$. Within our GiBUU framework, this integral is solved numerically using a Monte Carlo method.

In the following, we first present our results for electron and neutrino induced reactions and compare to other models.

\subsection{Inclusive electron scattering}

 Electron scattering off nuclei in the regime of energy transfers between $0.1$ and $1\GeV^2$ has been addressed by several experiments within the last two decades, for a recent review cf.~\refcite{Benhar:2006wy}. Comparing the measured nuclear cross sections to the nucleon cross sections, several modifications could be observed. First of all, nuclear Fermi motion leads to a smearing of the peak structures such in the quasielastic and  $\Delta$ regions. Furthermore, one observes a quenching of the spectral strength around the quasielastic peak, which has also been interpreted as a violation of the Coulomb sum-rule~\cite{baran1988,Day:1993md,ZGHICHE1994}. In contrast to the quenching in the peak region one observed an enhancement in the so-called \textit{dip-region} in between quasielastic and $\Delta$ peak. The peak position of the $\Delta$ resonance was found to be both $A$ and $Q^2$ dependent~\cite{Barreau1983,Sealock:1989nx}: a shift towards lower masses for $Q^2\lesssim 0.1$ and a shift towards higher masses for higher $Q^2$~\cite{OConnell1984,Sealock:1989nx,Chen:1990kq,Anghinolfi:1996vm}.
\subsubsection{Scattering of Oxygen}

\begin{figure}
   \centering
  \includegraphics[scale=\singleplotscale]{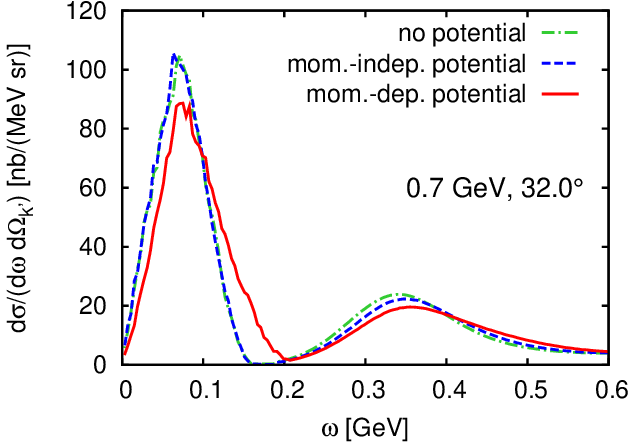}
 \caption{(Color online) The inclusive electron cross section $\dd\sigma / (\dd \omega \dd\Omega_{\kpr})$ on $^{16}$O  as a function of the energy transfer $\omega$ for a beam energy of 0.7 GeV and a scattering angle of $\theta_{\myvec{k}'}=32^\circ$. The plot shows the results for different nucleon potentials: no potential (dash-dotted line), momentum-independent potential  (dashed line) and momentum-dependent potential (solid line). The calculations do not include in-medium changes of the widths.}
\label{fig:electron_EQS}
\end{figure}

Here the results of our model are compared to data and the impact of the most prominent model ingredients is investigated.
In \fig{\ref{fig:electron_EQS}} we show our results for the inclusive reaction $^{16}$O$\left(e^-,e^-\right)X$ at a beam energy of 700 MeV and for different nucleon mean field potentials, in-medium changes to the width have been neglected. The dash-dotted curve denotes the result without potentials, including only Fermi motion and Pauli blocking. The presence of a momentum-independent potential (dashed curve) does not practically change the QE peak ($\omega=0.-0.15 \GeV$). However, the single-pion region ($\omega\gtrsim 0.2 \GeV$) is modified. This is due to the effect that the $\Delta$ --- dominating this region --- is less strongly bound than the nucleon. As a consequence, more energy must be transferred to the nucleon
to compensate the binding. When the momentum-dependent mean field is included (solid curve), then the faster (on average) final state nucleons experience a shallower potential than the initial state ones. Also the resonance potential gets shallower for higher momentum. Therefore, even more energy must be transferred by the photon leading to a broadening of the QE peak towards a higher energy transfer $\omega$ and to a shift of approximately $8\MeV$; also the single pion spectrum is slightly shifted towards higher energies and broadened.
A similar result has also been obtained within the Walecka model~\cite{Rosenfelder:1980nd}.  There the nucleon mass acquires an effective mass $M^\star(\myvec{r})$ such that the nucleon energy is given by
$
E_{\myvec{p}}=\sqrt{\myvec{p}\,^2+M^\star(\myvec{r})}
$ 
which can be rewritten as
$
E_{\myvec{p}}=\sqrt{\myvec{p}\,^2+M} +V(\myvec{r},\myvec{p})
$ 
with the momentum-dependent potential
\[
V(\myvec{r},\myvec{p})=\sqrt{\myvec{p}\,^2+M^\star(\myvec{r})}-\sqrt{\myvec{p}\,^2+M} \; .
\]
For small momenta ($|\myvec{p}|\ll M^\star$, $|\myvec{p}|\ll M$), one obtains
a simple quadratic dependence of the potential on the momentum
\[
V(\myvec{r},\myvec{p})\approx p^2 \frac{M-M^\star(\myvec{r})}{2MM^\star(\myvec{r})}+M^\star(\myvec{r})-M \ePunkt
\]
Rosenfelder~\cite{Rosenfelder:1980nd} suggests that the value of $M^\star$ can then be used to fit the QE peak. We emphasize however, that we do not fit our potential to the electron data since it has already been fixed  by nucleon-nucleus scattering~\cite{Teis:1996kx}.

\begin{figure}
   \centering
  \includegraphics[scale=\singleplotscale]{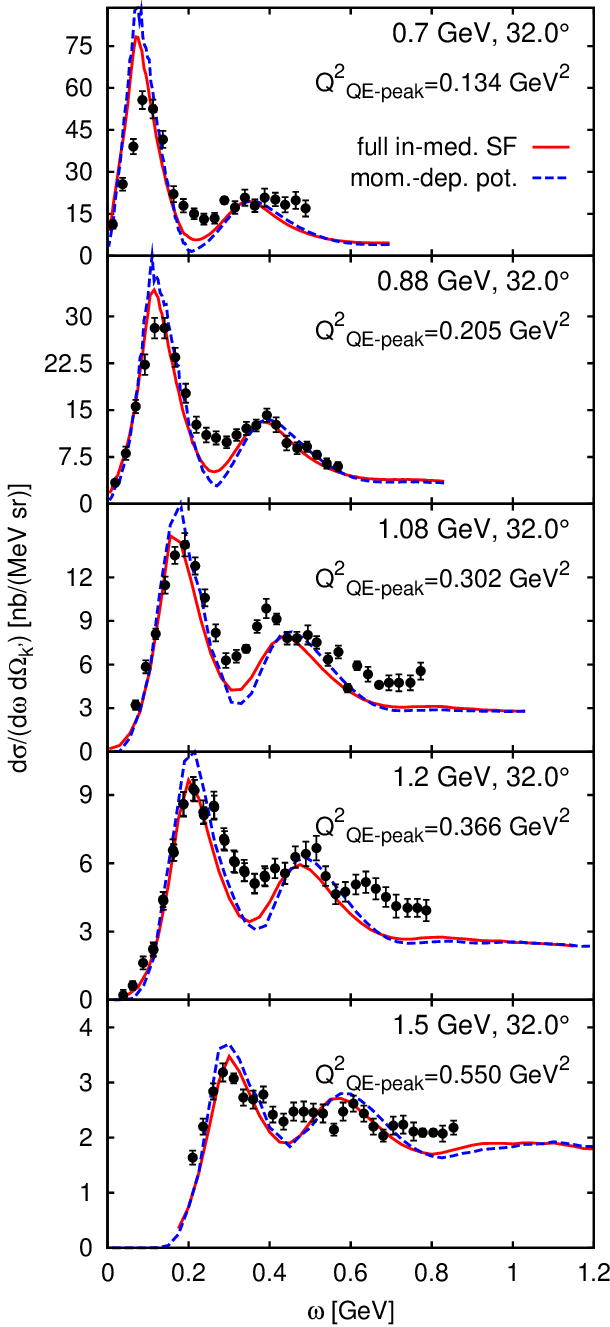}
\caption{(Color online) The inclusive electron cross section $\dd\sigma / (\dd \omega \dd\Omega_{\kpr})$ on $^{16}$O  as a function of the energy transfer $\omega$ at five distinct fixed electron energies (0.7, 0.88, 1.08, 1.2 and 1.5 GeV) and a scattering angle of $\theta_{\myvec{k}'}=32^\circ$. The dashed lines denote our result, where we include all in-medium modifications besides collisional broadening.  The solid lines denote the full calculation, which includes in-medium changes of the width. The data are taken from~\refscite{Anghinolfi:1996vm,Anghinolfi:1995bz} and the parameter $Q^2_\text{QE-peak}$ is evaluated according to \refeq{eq:q2_qePeak}.}
\label{fig:electronAll}
\end{figure}

\reffig{fig:electronAll} shows the comparison of our model to the data measured by Anghinolfi \refetal{Anghinolfi:1996vm,Anghinolfi:1995bz}. The overall agreement with data is good, specially when the in-medium widths are taken into account. In particular, the description of the QE peak is successful except at the lowest beam energy (0.7~GeV). At large energy transfers, the lack of double pion production strength leads to an underestimation of the data

\begin{figure}
\centering
\includegraphics[scale=\singleplotscale]{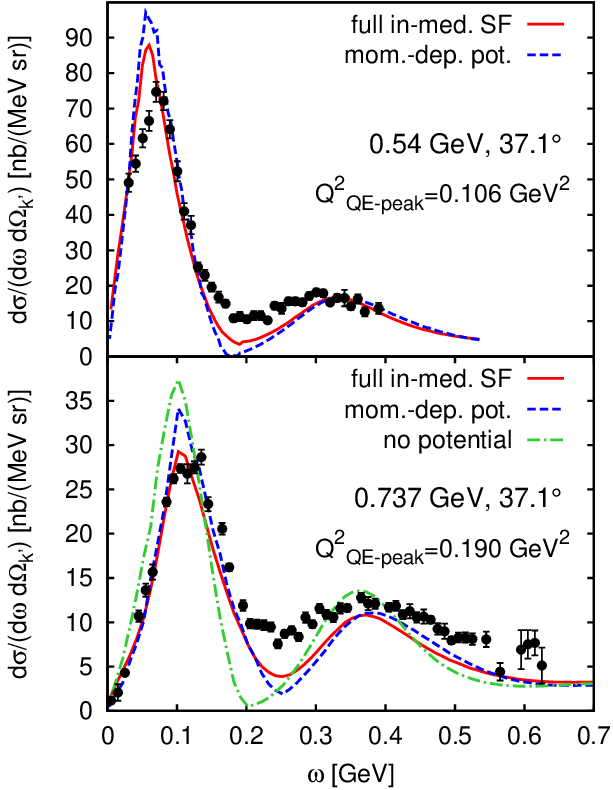}
\caption{(Color online) The inclusive electron cross section  $\dd\sigma / (\dd \omega \dd\Omega_{\kpr})$ on $^{16}$O  as a function of the energy transfer $\omega$ for a beam energy of 0.537 GeV and 0.737 GeV and a scattering angle of $\theta_{\myvec{k}'}=37.1^\circ$ in comparison to the data measured by O'Connell \refetal{OConnell1984,OConnell:1987ag}.  The parameter $Q^2_\text{QE-peak}$ is evaluated according to \refeq{eq:q2_qePeak}.}
\label{fig:oxygen_oConnell}
\end{figure}

To analyze the problem at $E_\text{beam}=0.7\GeV$ more closely, we compare in \reffig{fig:oxygen_oConnell} our model to measurements performed by O'Connell \refetal{OConnell1984,OConnell:1987ag}, where the kinematical constraints are similar to the ($E_\text{beam}=0.700 \GeV$, $\theta_{\myvec{k}'}=32.0^\circ$)-run performed by Anghinolfi {\etal} To compare experiments one often defines the so-called $Q^2$ at the QE-peak. It corresponds to the $Q^2$ at which the center-of-mass energy at the hadronic vertex equals the nucleon mass if one assumes a free nucleon target at rest. This parameter depends only on
the beam energy $E_\text{beam}$ and the electron scattering angle $\theta_{\myvec{k}'}$
and is given by
\eb
Q_\text{QE-peak}^2=2{M_N} \frac{E_\text{beam}^2(1-\cos\theta_{\myvec{k}'})}{{M_N}+E_\text{beam}(1-\cos\theta_{\myvec{k}'})} \ePunkt
\label{eq:q2_qePeak}
\ee
Note that it does not include any in--medium input, but it gives a simple estimate of the  $Q^2$  value at which the QE maximum is reached. As for the Anghinolfi data, we get also for the O'Connell data~\cite{OConnell1984,OConnell:1987ag} good correspondence for the measurement with higher $Q^2_\text{QE-peak}$ of $0.190\GeV^2$ while we overshoot the QE-peak height for the lower $Q^2_\text{QE-peak}=0.106\GeV^2$.

\begin{figure}[h!]
\centering
\includegraphics[scale=\singleplotscale]{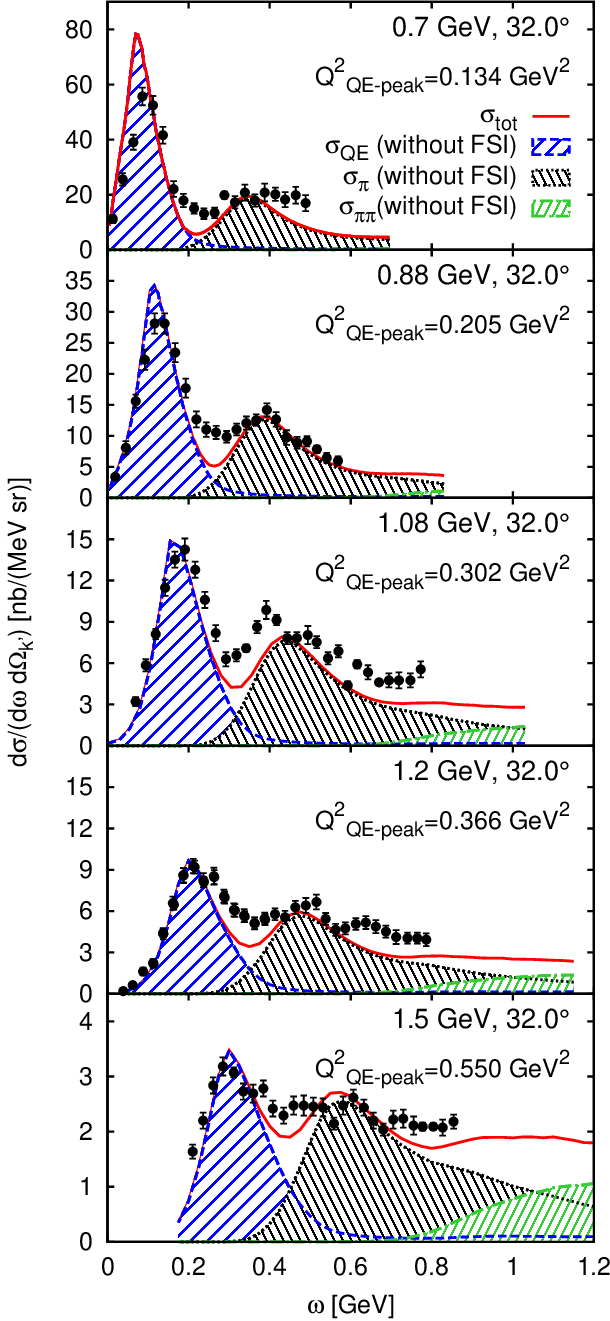}
\caption{(Color online) The inclusive electron cross section $\dd\sigma / (\dd \omega \dd\Omega_{\kpr})$ on $^{16}$O  as a function of the energy transfer $\omega$ at five distinct fixed electron energies (0.7, 0.88, 1.08, 1.2 and 1.5 GeV) and a scattering angle of $\theta_{\myvec{k}'}=32^\circ$. The solid lines denote our full result, where we include all in-medium modifications and in particular in-medium changes of the width. The data are taken from~\cite{Anghinolfi:1996vm,Anghinolfi:1995bz}. The dashed lines show the quasielastic contribution, the dotted ones the single-$\pi$ and the dash-dotted ones the $2\pi$ contribution to the initial scattering process. This result does not include any FSI of the outgoing particles. The parameter $Q^2_\text{QE-peak}$ is evaluated according to \refeq{eq:q2_qePeak}.}
\label{fig:electronAll_channels}
\end{figure}

In \reffig{fig:electronAll_channels}, we show the contribution of the different production mechanisms to the total electron-nucleus cross section  including all in-medium modifications. The quasielastic, single-$\pi$ and $2\pi$ contributions to the initial scattering process are shown. One observes that when going from low to high beam energies, the importance of single-$\pi$ and $2\pi$ production mechanism gradually increases, whereas at low energies the quasielastic contribution dominates. Note that this result does not include any FSI of the outgoing particles except for the mean-field potential and of the outgoing nucleon and the classification into different channels is solely based on the initial vertex and not on the final-state multiplicities.

\subsubsection{Impact of the initial phase space distribution.} In the following we compare results for different phase space distributions of the target nucleons. As outlined in  \refsec{sec:groundstate}, we assume for the nuclear ground state that the positions of the nucleons are distributed according to density parametrizations obtained from low-energy electron scattering and Hartree-Fock calculations. The momenta of the nucleons are distributed according to a local Thomas-Fermi (LTF) approximation, i.e.~it is assumed that at each space-point $\myvec{r}$ the nucleon momenta  occupy a uniform sphere in momentum space with a radius given by the Fermi momentum $p_f(\myvec{r})$. In \refsec{sec:momDensities} we have already compared the simpler global Fermi Gas  model to LTF finding quite different  momentum densities.

For a nucleon at rest, QE scattering takes place at a given energy transfer (for fixed incoming beam energy and scattering angle)

\eb
\omega=\frac{Q^2_\text{QE-peak}}{2M}=\frac{E_\text{beam}^2(1-\cos\theta_{\myvec{k}'})}{E_\text{beam}(1-\cos\theta_{\myvec{k}'})+M} \,.
\ee
The finite target nucleon momenta within a Fermi gas lead to a finite range of possible $\omega$ centered roughly around $\omega=Q_\text{QE-peak}^2/2M$, for which a QE-event can be realized. The size of this range is determined by the Fermi momentum: the larger the Fermi momentum the broader $\omega$ range is present.

To compare the LTF and GFG distributions we use the very same physics input, in particular the same potentials and widths, varying only the initial $\myvec{r}$ and $\myvec{p}$ distributions focussing on the quasielastic peak.

\begin{figure}
   \centering
  \includegraphics[scale=\singleplotscale]{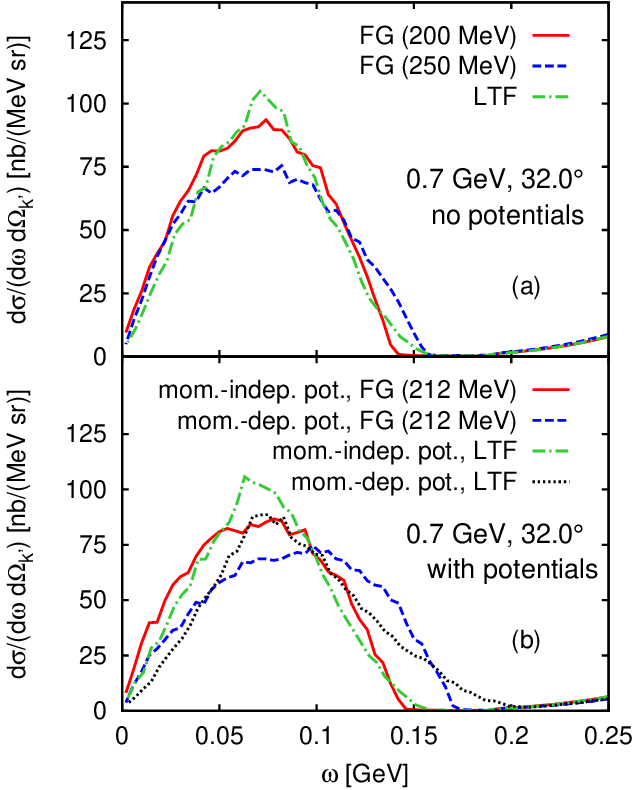}
 \caption{(Color online) The inclusive electron cross section $\dd\sigma / (\dd \omega \dd\Omega_{\kpr})$ on $^{16}$O  as a function of the energy transfer $\omega$ for a beam energy of 0.7 GeV and a scattering angle of $\theta_{\myvec{k}'}=32^\circ$. The graphs in panel (a) show the results for a calculation where we neglected potentials and in-medium width modifications. The results for various assumptions concerning the momentum distribution of the target nucleons are shown: Fermi gas with Fermi momentum $p_f=0.2 \GeV$ (solid line), Fermi gas with Fermi momentum $p_f=0.25 \GeV$ (dashed line), momentum distribution according to local Thomas-Fermi (LTF) approximation (dash-dotted line). The calculations shown in panel (b) include mean field potentials, but also no in-medium changes of the widths. The GFG result in panel (b) was obtained with a Fermi momentum of $212\MeV$.}
\label{fig:electron_FG}
\end{figure}
In panel (a) of \reffig{fig:electron_FG}, we show the results for different input distributions neglecting all in-medium modifications besides Pauli blocking and Fermi motion. The dashed-dotted curve represents the result with our standard momentum distribution according to the local-Thomas-Fermi ansatz. One observes that an initial distribution according to the GFG with a constant Fermi momentum of $0.2\GeV$ yields already a slightly lower and broader QE-peak compared to the LTF result. Fig.~\ref{momDist} shows the momentum densities of the target nucleons for the LTF scheme (solid line) and the Fermi gas with $p_f=200\MeV$. Note that both distributions have the same average for the absolute value of the nucleon momentum ($150\MeV$). However, within the LTF ansatz there are more nucleons with low momenta which leads to more strength around $\omega\approx Q_\text{QE-peak}^2/2M$. The surplus of high-momentum  nucleons for the LTF case  compared to the GFG leads to a slightly higher maximal photon energy for LTF and to a slight broadening of the peak at low and high photon energies for LTF.
With a further increase of the Fermi momentum to $0.25\GeV$ \footnote{Note that a Fermi momentum of $0.25\GeV$ is not very realistic: a realistic density profile for the dilute Oxygen nucleus~\cite{osetPionicAtoms,DeJager:1974dg} gives within LTF only a density-averaged Fermi momentum of ca.~$200\MeV$.} we obtain a further broadening of the QE peak and a prominent reduction of the peak height. The peak position is however insensitive to the magnitude of the Fermi momentum, so that for a GFG the height of the QE-peak could be fitted by a variation the Fermi momentum parameter.

In panel (b) of \reffig{fig:electron_FG}, we additionally consider the impact of two different mean-field potentials for the nucleons and compare the LTF results to GFG results with $p_f=212 \MeV$ (this Fermi momentum gives a  nuclear density of $0.84 \fm^{-3}$ which is approximately the same as the average nuclear density of an {\oxygen} nucleus). In a  GFG, the density is constant and, therefore, all nucleons are bound by a similar potential which only differs due to its momentum-dependence.In  the LTF description, however, the nucleons feel quite different potentials depending on their position since the nuclear density becomes position dependent. The solid and dash-dotted curves in panel (b) of \reffig{fig:electron_FG} show the results for a momentum-independent potential. The comparison of  our results with such a potential to those without potential (panel (a)) shows that a momentum independent potential has almost no visible impact on the results because the potentials for the incoming and outgoing nucleons are the same and the energy transfer stays constant. Also the slightly lower nucleon in-medium masses  due to the mean-field potential do not lead to a sizable modification of the cross section.

The dashed and dashed-dotted curves on panel (b) of \reffig{fig:electron_FG} show the results for the GFG and LTF distributions when including a momentum dependent potential. One observes in both cases a shift of strength towards higher photon energies. This shift comes from the potential difference of the slow and strongly bound target nucleon to the faster and less strongly bound final-state nucleon. However, there is a qualitative difference in the spectra for LTF and GFG: the QE-peak for the GFG  is shifted by roughly $25 \MeV$ towards higher $\omega$ while it hardly shifts for the LTF ansatz. This feature is caused by the fact that the contribution of the slow target nucleons dominates the region around the in-medium QE-peak. Within the LTF scheme, these slow target nucleons tend to sit at  low densities (cf. \reffig{momDensDist}) and, therefore, the region around the QE-peak is hardly affected by the medium. For the GFG, however, such correlation of density and momentum is absent and, both peak and off-peak strength are shifted towards higher masses.

This analysis has shown an important difference in between the global Fermi gas  and the local Thomas-Fermi approximation. We emphasize that in the LTF case, in contrast to the GFG, nucleon position and nucleon momentum are correlated. If one uses  the same momentum-dependent potential in both cases, the peak positions of the quasielastic peak and also of the resonance peaks barely shift for the LTF distribution while there are large shifts in the GFG.


\subsection{Inclusive neutrino scattering}
\begin{figure}
  \centerline{\includegraphics[scale=\singleplotscale]{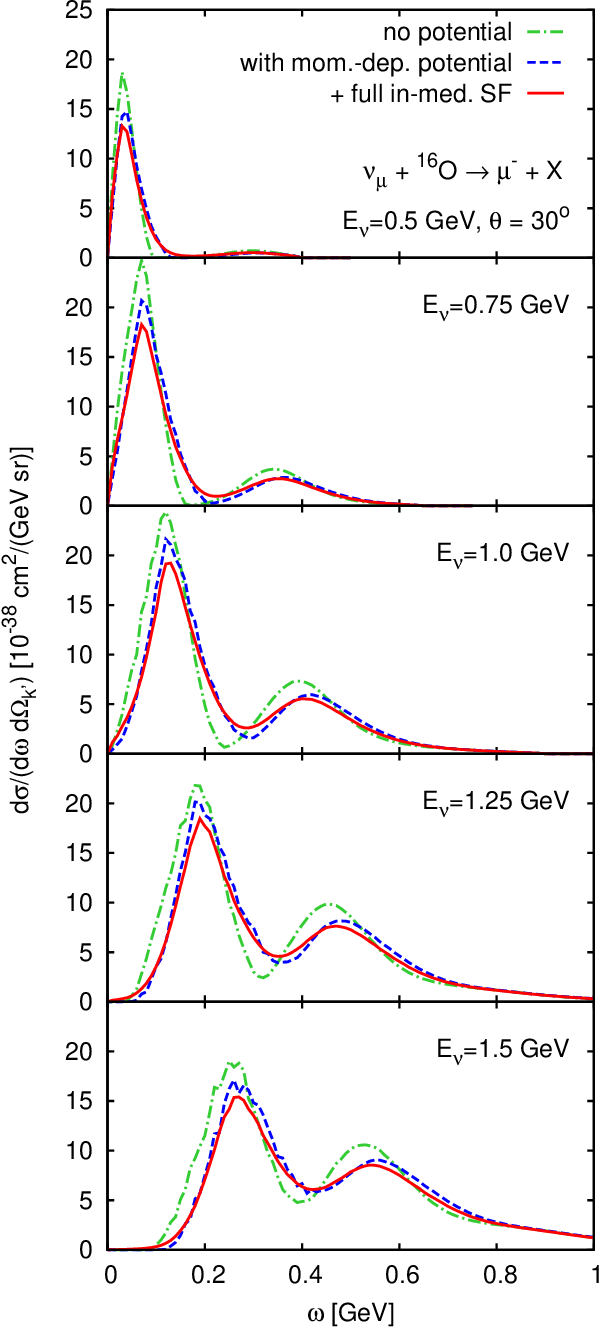}}
  \caption{(Color online) Inclusive CC neutrino cross section $\dd\sigma / (\dd \omega \dd\Omega_{\kpr})$ on $^{16}$O  as a function of the energy transfer $\omega$ at five distinct fixed neutrino energies (0.5, 0.75, 1.0, 1.25 and 1.5 GeV) and a scattering angle of $\theta_{\myvec{k}'}=30^\circ$. The dash-dotted lines denote our calculation without any mean-field potential while a momentum-dependent potential is included in the calculation shown with the dashed lines. The solid lines denote the full calculation, which includes in addition in-medium changes of the width.   \label{fig:neutrino_CC_inclusive}}
\end{figure}
\begin{figure}
  \centerline{\includegraphics[scale=\singleplotscale]{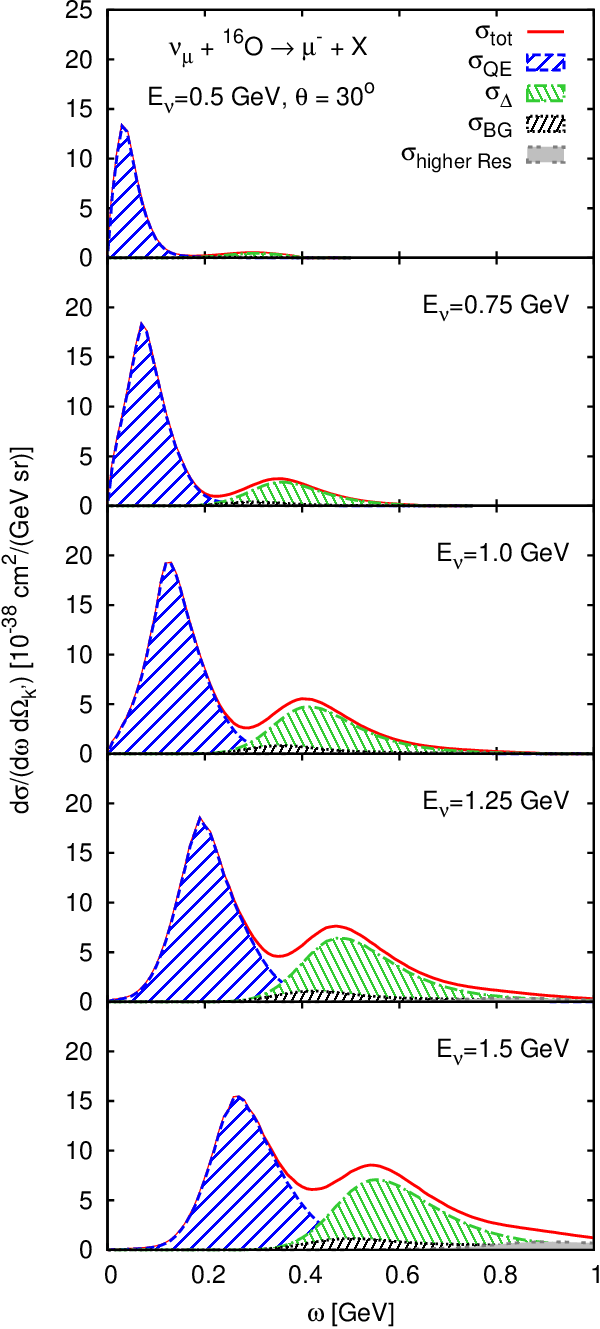}}
  \caption{(Color online) Inclusive CC neutrino cross section $\dd\sigma / (\dd \omega \dd\Omega_{\kpr})$ on $^{16}$O  as a function of the energy transfer $\omega$ at five distinct fixed neutrino energies (0.5, 0.75, 1.0, 1.25 and 1.5 GeV) and a scattering angle of $\theta_{\myvec{k}'}=30^\circ$. The solid line denotes our full result, where we include all in-medium modifications and in particular in-medium changes of the width. The dashed lines show the quasielastic contribution, the dotted ones the single-$\pi$ background and the dash-dotted ones the contribution coming from an initial $\Delta$ excitation. Higher resonance contributions are indicated by the double-dashed lines. \label{fig:neutrino_CC_inclusive_contribution}}
\end{figure}

\begin{figure}
  \centerline{\includegraphics[scale=\singleplotscale]{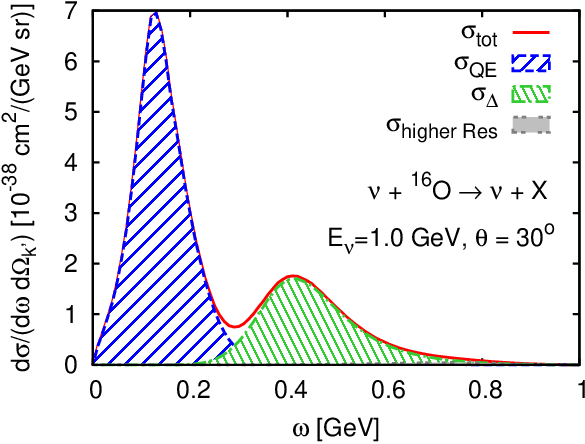}}
  \caption{(Color online) Inclusive NC neutrino cross section $\dd\sigma / (\dd \omega \dd\Omega_{\kpr})$ on $^{16}$O  as a function of the energy transfer $\omega$ at a distinct fixed neutrino energies of 1 GeV and a scattering angle of $\theta_{\myvec{k}'}=30^\circ$. The solid line denotes our full result, where we include all in-medium modifications and in particular in-medium changes of the width. The dashed lines show the quasielastic contribution, the dash-dotted ones the contribution coming from an initial $\Delta$ excitation and the double-dashed ones the contributions from higher resonances.  \label{fig:neutrino_NC_inclusive}}
\end{figure}
In \reffig{fig:neutrino_CC_inclusive} we show results for CC $\nu_\mu$ scattering off oxygen. The dash-dotted lines show our result without any mean-field potential while it is included in the calculation denoted with the dashed line. The solid line includes, in addition, the in-medium spectral function introduced in \refeq{eq:inmedspecfunc}, which accounts for the collisional broadening of the baryons in the medium. The effects of these in-medium corrections are identical to the electron scattering case shown in \reffig{fig:electron_EQS} and \reffig{fig:electronAll}: We find a broadening and a shift of the QE and resonace peaks as a consequence of the momentum-dependent potential and the in-medium width. Thus, the good description of the inclusive electron scattering serves as a direct benchmark for our neutrino calculations even though there are still no neutrino data to compare with.

First results for CC inclusive neutrino scattering off nuclei were already presented in \refscite{Leitner:2006ww,Buss:2007ar}. There, only the excitation of the $\Delta$ resonance was taken into account --- in this work, we improved on this and include 13 resonances and a non-resonant pion background. The strength of those contributions can be seen in \reffig{fig:neutrino_CC_inclusive_contribution}. The left peaks have their origin in QE scattering (compare the total yield given by the solid lines to the QE yield given by the dashed ones). With increasing $\omega$, one enters the pion-production region, which is dominated by initial $\Delta$ excitation (solid vs.~dash-dotted lines). The non-resonant single pion background also contributes in this region (dotted lines) while the impact of the higher resonances is only visible at high beam energies (double dashed lines in lowest panel with $E_\nu=1.5\GeV$). \reffig{fig:neutrino_NC_inclusive} shows the corresponding plot for NC scattering off Oxygen.

\subsection{Comparison to other models}

Finally, we wish to compare our results to the ones obtained in other models, focusing on the results obtained for Oxygen nuclei.
The work of Butkevich \refetal{Butkevich:2007gm,Butkevich:2005ph} addresses the quasielastic peak region in a relativistic distorted wave impulse approximation (RDWIA) approach. Their latest model for the ground state spectral function includes both a $75 \%$ shell-model and a $25 \%$ high-momentum contribution. When calculating inclusive cross sections, the final state wave functions are obtained using a real optical potential. This procedure neglects the broadening of the outgoing nucleon in the medium but incorporates a shift of the outgoing nucleon energy.
%
A direct comparison to models based on DWIA with \emph{complex optical potentials} (\emph{RDWIA exclusive} result in Butkevich \refetal{Butkevich:2007gm} or, e.g., Martinez \refetal{Martinez:2005xe}, Maieron \refetal{Maieron:2003df}) is not possible since the imaginary part leads to a flux reduction in a particular channel. Thus, those models rather describe exclusive quasifree single-nucleon knockout processes than fully inclusive scattering which we consider in this work.

Meucci \refetal{Meucci:2003cv,Meucci:2003uy} apply a relativistic Green's function approach to both inclusive and exclusive processes. They achieve an impressive description for the data on both the longitudinal and transverse response functions in the QE region. In the Green's function framework a complex optical potential can be incorporated when calculating inclusive cross sections without having the DWIA-problem of flux reduction.

The model of Benhar, Nakamura and collaborators \cite{Benhar:2005dj,Benhar:2006qv,Benhar:2006nr,Nakamura:2007pj} is based on non-relativistic nuclear many body theory and impulse approximation. It includes realistic spectral functions for the hole states obtained from $(e,e'p)$ data combined with theoretical nuclear matter calculations using the local density approximation. For the final state spectral function, they rely on a correlated Glauber approximation which leads to an energy shift of the cross section and to a redistribution of the strength (quenching of the peak and enhancement of the tail).

The framework of Benhar {\it et al.} provides a state-of-the-art description for the hole spectral functions including both real and imaginary parts of the self energy. In our model, only the real part of the hole self energy is taken into account for the initial nucleons (through the mean-field potential)  while the imaginary part is neglected.

The model of Gil, Nieves and others, applicable to both inclusive electron~\cite{Gil:1997bm} and neutrino reactions~\cite{Nieves:2004wx}, takes into account nuclear corrections beyond Pauli blocking: They include long and short range correlations as well as particle and hole spectral functions. As in our approach, they neglect the imaginary part of the self energy in the hole spectral function. Within their model Gil \refetal{Gil:1997bm} achieve very good agreement with data for electron-nucleus scattering.
Nuclear correlations are also taken into account by Singh, Oset and collaborators \cite{Singh:2006ci,Singh:1992dc} for neutrino scattering. They renormalize the weak transition strength and are found to be very important at low momentum transfers.


\begin{figure}
  \centerline{\includegraphics[scale=\singleplotscale]{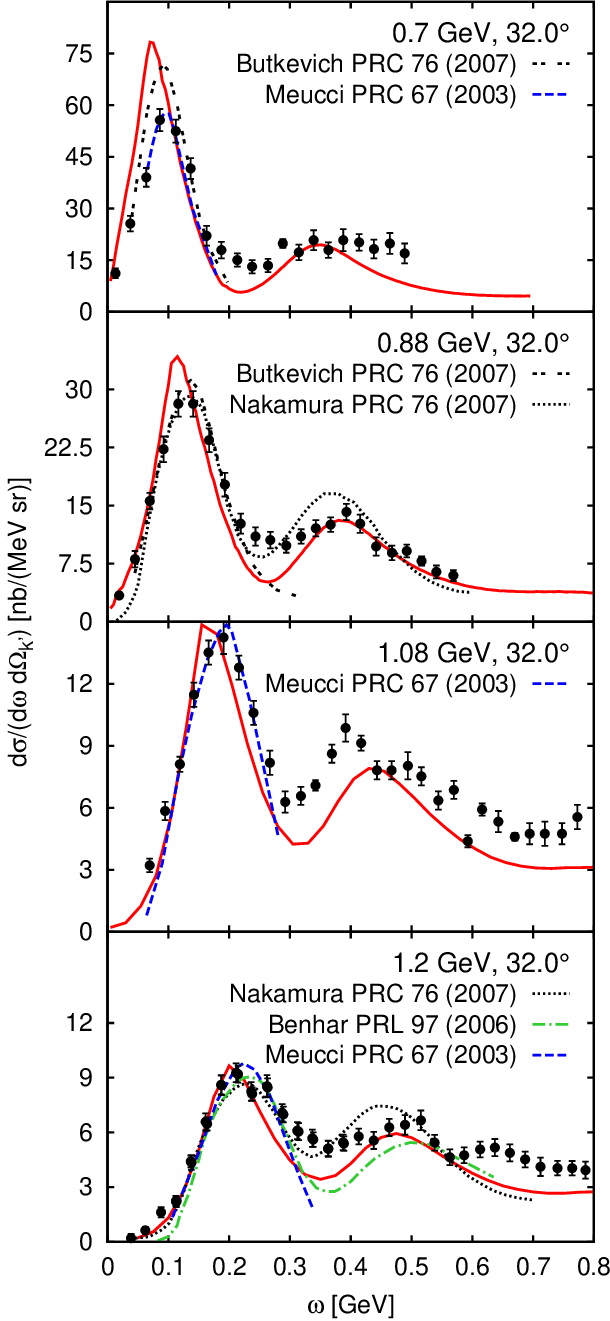}}
  \caption{(Color online) The inclusive electron cross section $\dd\sigma / (\dd \omega \dd\Omega_{\kpr})$ on $^{16}$O  as a function of the energy transfer $\omega$ at four distinct fixed electron energies (0.7, 0.88, 1.08 and 1.2) and a scattering angle of $\theta_{\myvec{k}'}=32^\circ$. The solid lines denote our full result, the dashed result is taken from Meucci \refetal{Meucci:2003cv}, the double dashed one from Butkevich \refetal{Butkevich:2007gm} (RDWIA), the dotted one from Nakamura \refetal{Nakamura:2007pj} and the dashed-dotted one from Benhar \refetal{Benhar:2006qv}. The data are taken from~\cite{Anghinolfi:1996vm,Anghinolfi:1995bz}.}
  \label{electronVergleichModelle}
\end{figure}
\reffig{electronVergleichModelle} shows a comparison of our result including in-medium spectral functions (solid line) for inclusive electron scattering off {\oxygen} to the results of other models.
The dotted curves show the latest results of Nakamura, Benhar \refetal{Nakamura:2007pj}. Their model gives a better description of the QE-peak than ours at $880\MeV$. However, at $1200\MeV$ both achieve equally good descriptions of the data in the QE peak. In the pion-production region,
Nakamura tends to overshoot the data and the $\Delta$-peak position seems somewhat low in $\omega$. The data are also underestimated in the model of \refcite{Benhar:2006qv} (dash-dotted line). Our model lacks strength in the dip region but describes properly the magnitude and position of the $\Delta$ peak.
The framework of Butkevich \refetal{Butkevich:2007gm}, which only includes QE scattering, fails in the same kinematical situation as our model ($700\MeV$, double-dashed line in upper panel) but leads to very good results at $880\MeV$. The Green's function approach of Meucci \refetal{Meucci:2003cv} is able to describe very well all the data in the QE region.

\begin{figure}
  \centerline{\includegraphics[scale=\singleplotscale]{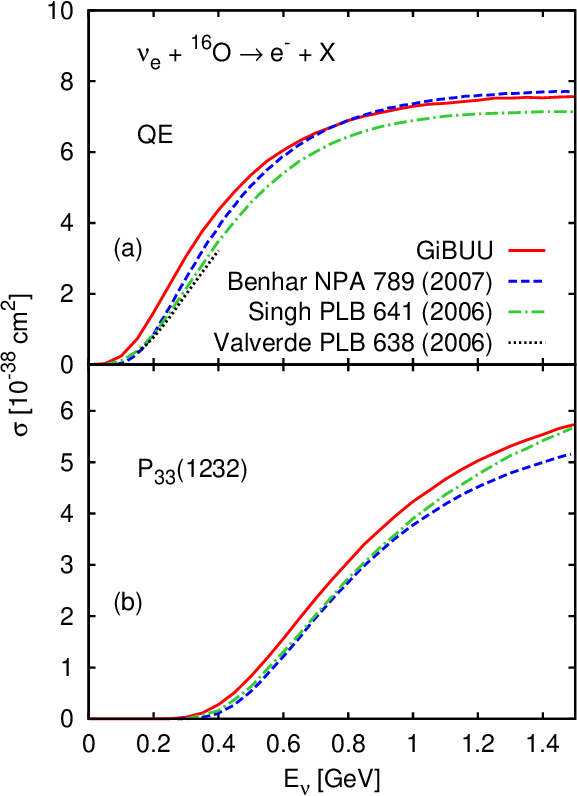}}
  \caption{(Color online) Inclusive CC electron neutrino cross section $\sigma$ on $^{16}$O as a function of the neutrino energy and integrated over the leptonic variables. (a) shows the QE contribution, (b) the $\Delta$ contribution. Our results are denoted by the solid line. We compare to Benhar \refetal{Benhar:2006nr} (their ``SF-PB'' calculation), Singh \refetal{Singh:2006ci} and Valverde, Nieves and others~\cite{Valverde:2006zn} (their full model).  \label{fig:neutrino_CC_inclusive_comparison}}
\end{figure}

In \reffig{fig:neutrino_CC_inclusive_comparison} we compare our CC calculation to the models introduced above for the integrated inclusive cross section on Oxygen as a function of the neutrino energy. Panel (a) shows the contribution from QE scattering while panel (b) shows the pure $\Delta$ contribution. Our full result is denoted by the solid line labeled ``GiBUU''. The overall agreement with the other models is satisfactory. Focussing on the QE contribution, our curve is higher than other calculations at lower neutrino energies. Also, for the $\Delta$ our calculation is slightly higher. However, note that differences are already present at the nucleon level and do not only arise from the different treatment of nuclear corrections.

Comparing to electron data, we conclude that our model is able to describe the QE region for beam energies above $~1\GeV$ while it fails for lower beam energies where nuclear many-body effects become important and impulse approximation breaks down \cite{Ankowski:2008df}. From the good agreement at higher energies at the QE peak and also in the single-$\pi$ region, we can conclude, that our low-density ansatz for the in-medium width (cf.~\refeq{inMed_Imag}) and the inclusion of a proper potential incorporate the main nuclear corrections.

\section{Conclusions and Outlook}

In this work, we have presented a model for elementary lepton-nucleon scattering and inclusive scattering off nuclei via EM, CC and NC at intermediate lepton energies, i.e., beam energies ranging from 0.5 to 2 GeV. In particular, the model for the elementary vertex is improved considerably compared to our earlier work in \refcite{Buss:2007ar}.

In this energy regime, the scattering process is dominated by three contributions: quasielastic scattering, resonance excitation and non-resonant pion production.
QE scattering and resonance excitation (we include 13 resonances) are both described with a relativistic formalism that incorporates recent form factor parametrizations. EM, CC and NC form factors are connected via isospin relations. For the nucleon vector form factors we apply the latest BBBA-2007 analysis accounting for new electron scattering data; the resonance vector form factors are based on the recent MAID analysis for the helicity amplitudes. The axial couplings are obtained applying the PCAC theorem. For the $\Delta$ resonance, the axial form factor was refitted to the ANL data. For the non-resonant pion background, we have used a technique based on invariant amplitudes taken from MAID, allowing us to incorporate background terms not only for EM, but also for CC processes, where the additional non-vector parts were fitted to the ANL data for total pion production cross sections.

When scattering leptons off bound nucleons, in-medium corrections have to be considered. We use a local Thomas-Fermi approximation for the phase space density of the bound nucleons based on realistic nuclear densities. Pauli blocking is naturally included in this way. In the initial QE scattering or resonance excitation process, we take into account full in-medium kinematics, momentum-dependent mean-field potentials and the in-medium spectral functions of the outgoing baryons. The parameters of the mean-field potential were independently fixed with nucleon-nucleus scattering data. The imaginary part of the self energies entering the spectral functions are calculated in a consistent way employing the low-density approximation. The correct normalization of the spectral functions is ensured by extracting the real part of the self energies from dispersion relations.

Good agreement to the experimental electron scattering data on Oxygen at beam energies ranging from $0.7 - 1.5 \GeV$ is achieved both in the QE and in the pion production region. The overall agreement is improved by taking into account mean fields and in-medium spectral functions in addition to a local Fermi gas momentum distribution. The momentum-dependence of the mean field reshapes the QE peak considerably due to a target-momentum dependent energy loss and, in addition, shuffles strength into the dip region, which is conventionally attributed to 2N excitations; also here the description is considerably improved due to the collisional broadening of the QE peak. Good description of data is also achieved in the single-pion region, while at higher beam energies, the data are underestimated at high photon energies $\omega$ due to the fact that non-resonant $2\pi$-production channels have not yet been included.

Still unsolved is the discrepancy at the QE peak at very low $Q^2$ of $Q^2 \le 0.15$ where our model overestimates both the Anghinolfi~\cite{Anghinolfi:1996vm,Anghinolfi:1995bz} and O'Connell~\cite{OConnell1984,OConnell:1987ag} data and does not resemble the peak shift. In view of this, we studied the influence of the initial momentum distribution of the nucleons. We compared our local Thomas-Fermi approach to the widely used global Fermi gas approximation where the Fermi momentum does not depend on the density, and thus, nucleon momentum and position are not correlated as in the local Thomas-Fermi ansatz. Qualitative and quantitative difference have been found, in particular the shift of the QE peak is easily obtained in the global Fermi gas picture, while the peak hardly shifts in the more realistic local Thomas-Fermi picture.

Taking the in-medium modifications on the electron scattering results as a benchmark, we have made predictions for CC and NC inclusive scattering of Oxygen at beam energies ranging from 0.5 up to 1.5 GeV.

To summarize, we have presented in this article a consistent model for EM, CC and NC reactions off nucleons and inclusive scattering off nuclei in the energy regime of interest for current neutrino oscillation experiments. In-medium corrections are found to be of considerable importance and required for a good description of electron scattering data, and, therefore, also for neutrino induced reactions.
The extension of the  present model to semi-inclusive processes can be performed with the GiBUU transport model and results will be given elsewhere.

\begin{acknowledgments}
We thank L.~Tiator (Mainz) for his kind support and for providing us a compilation of the MAID amplitudes and O.~Lalakulich for many discussions.
We also thank the GiBUU team for instructive discussions. LAR acknowledges financial support from the Seneca Foundation. This work was supported by the Deutsche Forschungsgemeinschaft.

\end{acknowledgments}

\appendix
\section{Symmetries, currents and isospin relations}

\subsection{Hadronic currents}
At quark level, one can directly relate the electromagnetic and the weak vector currents assuming only isospin symmetry of the strong interactions. If isospin symmetry is a good symmetry of the particular hadronic system, one expects the obtained relations to be independent of the details of the hadronic structure.
We then find the following structure for the electromagnetic current
\begin{equation}
J^{\mu}_{\subEM}= \frac{1}{2} V^{\mu}_Y + V^{\mu}_3 \ePunkt   \label{eq:vectorcurrentEM}
\end{equation}
$V^{\mu}_3$ is the third component of the isovector (isospin) current and $V^{\mu}_Y$ the isoscalar (hypercharge) current.

The vector part of the charged current is given by
\begin{equation}
   V^{\mu}_{\subCC}    = V^{\mu}_1 + \ii V^{\mu}_2\eKomma  \label{eq:vectorcurrentCC}
\end{equation}
and the vector part of the neutral current by
\begin{equation}
V^{\mu}_{\subNC}=(1-2\, \sinwein) V^{\mu}_3 - \sinwein V^{\mu}_Y - \frac{1}{2} V^{\mu}_S\eKomma   \label{eq:vectorcurrentNC}
\end{equation}
where $V^{\mu}_Y$ is the hypercharge,  $V^{\mu}_S$ is the strange (both isoscalar) and $V^{\mu}_{1,2,3}$ is the isovector current. We further assume that the matrix elements of the hadronic currents $V^{\mu}_{1,2}$ and $V^{\mu}_3$ are the same, being related by isospin rotation. This is known as ``conserved vector current hypothesis'' (CVC).

Also for the axial part, we assume the hadronic currents to have the same structure
\begin{equation}
A^{\mu}_{\subCC}= A^{\mu}_{1} + \ii A^{\mu}_{2}\eKomma \label{eq:axialcurrentCC}
\end{equation}
for charged currents and
\begin{equation}
A^{\mu}_{\subNC}= A^{\mu}_{3} + \frac12 A^{\mu}_S\eKomma  \label{eq:axialcurrentNC}
\end{equation}
for neutral currents which includes the isoscalar strange axial current $A^{\mu}_S$.
 $A^{\mu}_{1,2}$  and  $A^{\mu}_3$ are components of the same isospin vector.

\subsection{Isospin relations for QE scattering and resonance excitation}

\subsubsection{Isospin 1/2 \texorpdfstring{$\to$}{->} 1/2 transition}
\label{sec:iso12iso12}

In general, the current is given by
\begin{equation}
\myvec{J}^\mu=(J_1^\mu,J_2^\mu,J_3^\mu)=\myvec{V}^\mu-\myvec{A}^\mu\eKomma
\end{equation}
with
\begin{align}
\myvec{V}^\mu & = (V_1^\mu,V_2^\mu,V_3^\mu) = \mathcal{V}^\mu \, \frac{\myvec{\tau}}{2}\eKomma \\
\myvec{A}^\mu & = (A_1^\mu,A_2^\mu,A_3^\mu)= \mathcal{A}^\mu \, \frac{\myvec{\tau}}{2}\eKomma
\end{align}
where $\myvec{\tau}$ is the isospin 1/2 transition operator given by the Pauli matrices.

In a transition between isospin 1/2 states, both, isoscalar and isovector parts of the current contribute. Therefore, we further define for the hypercharge part
\begin{equation}
V^{\mu}_Y = \mathcal{V}_Y^\mu \unittwo\eKomma
\end{equation}
which enters \refeq{eq:vectorcurrentEM} and \refeq{eq:vectorcurrentNC}, and
\begin{align}
  V^\mu_S&=\mathcal{V}_S^\mu \unittwo\eKomma \\
  A^\mu_S&=\mathcal{A}_S^\mu \unittwo\eKomma
\end{align}
for the strange component entering \refeq{eq:vectorcurrentNC} and \refeq{eq:axialcurrentNC}, respectively. $\unittwo$ is the $2\times2$ unit matrix.

$\mathcal{V}^\mu$, $\mathcal{A}^\mu$, ... are given in \refeq{eq:QEVECcurrent} and \eqref{eq:QEAXcurrent} for QE, in \refeq{eq:vectorspinhalfcurrent} and \eqref{eq:axialspinhalfcurrent} for spin 1/2 and in \refeq{eq:vectorspinthreehalfcurrent} and \eqref{eq:axialspinthreehalfcurrent} for spin 3/2, \emph{but}, with the corresponding form factors: $F_i^V$ for $\mathcal{V}^\mu$, $F_i^s$ for $\mathcal{V}_S^\mu$, ...

Combining all that, we obtain for the electromagnetic transition matrix element
\begin{align}
\matrixelement{{N^*}^+}{J_{\subEM}^\mu}{p}&=\matrixelement{{N^*}^+}{V_3^\mu+\frac12 V^{\mu}_Y}{p} \nonumber \\
&=\matrixelement{{N^*}^+}{\mathcal{V}^\mu \frac{\tau_3}{2}+ \mathcal{V}^{\mu}_Y \frac{\unittwo}{2}}{p} \nonumber \\
&= \frac{\mathcal{V}^\mu + \mathcal{V}_Y^\mu}{2}\eKomma \nonumber \\
&\equiv \mathcal{V}^\mu_p\eKomma
\end{align}
where $\mathcal{V}^\mu_p$ incorporates the proton form factors $F_i^p$.
Analogously, one finds
\begin{align}
\matrixelement{{N^*}^0}{J_{\subEM}^\mu}{n}&= \frac{- \mathcal{V}^\mu + \mathcal{V}_Y^\mu}{2} \equiv \mathcal{V}^\mu_n\eKomma
\end{align}
where $\mathcal{V}^\mu_n$ incorporates the neutron form factors $F_i^n$. This yields the final relations
\begin{align}
  \mathcal{V}^\mu &=\mathcal{V}^\mu_p - \mathcal{V}^\mu_n\eKomma    \label{eq:vectorprotonminusneutron} \\
  \mathcal{V}_Y^\mu&=\mathcal{V}^\mu_p + \mathcal{V}^\mu_n\ePunkt    \label{eq:hyperprotonplusneutron}
\end{align}

The vector part of the charged current transition matrix element is
\begin{align}
  \matrixelement{{N^*}^+}{V_{\subCC}^\mu}{n}&= \matrixelement{{N^*}^+}{V^{\mu}_1 + \ii V^{\mu}_2}{n}\eKomma \nonumber \\
&=\matrixelement{{N^*}^+}{\mathcal{V}^{\mu} \tau_+}{n}\eKomma\nonumber \\
&=\mathcal{V}^\mu\eKomma
\end{align}
which together with \refeq{eq:vectorprotonminusneutron} relates EM and CC form factors.

Applying the same procedure to NC, one gets
\begin{align}
\matrixelement{{N^*}^+}{V_{\subNC}^\mu}{p} &=  (1-2\, \sinwein) \mathcal{V}^\mu  - \sinwein \mathcal{V}_Y^\mu  - \frac{1}{2} \mathcal{V}_S \eKomma \\
\matrixelement{{N^*}^0}{V_{\subNC}^\mu}{n} &=- (1-2\, \sinwein) \mathcal{V}^\mu  - \sinwein \mathcal{V}_Y^\mu  - \frac{1}{2} \mathcal{V}_S \ePunkt
\end{align}
Hereby, we relate the NC form factors via \refeq{eq:vectorprotonminusneutron} and \refeq{eq:hyperprotonplusneutron} to the EM proton and neutron form factors --- in addition, strange form factors coming from $\mathcal{V}_S$ have to be considered.

One can proceed in the same way for the axial part and finds
\begin{align}
   \matrixelement{{N^*}^+}{A_{\subCC}^\mu}{n}&=\matrixelement{{N^*}^+}{A^{\mu}_{1} + \ii A^{\mu}_{2}}{n}\eKomma\nonumber \\
&= \matrixelement{{N^*}^+}{\mathcal{A}^\mu \tau_+}{n}\eKomma \nonumber \\
&=\mathcal{A}^\mu\eKomma
\end{align}
while
\begin{align}
\matrixelement{{N^*}^+}{A_{\subNC}^\mu}{p}&=  \frac{\mathcal{A}^\mu + \mathcal{A}_S^\mu}{2} \eKomma \\
\matrixelement{{N^*}^0}{A_{\subNC}^\mu}{n}&=  \frac{-\mathcal{A}^\mu + \mathcal{A}_S^\mu}{2}\eKomma
\end{align}
which connects the NC axial form factors with the CC and the strange axial form factors.

\subsubsection{Isospin 1/2 \texorpdfstring{$\to$}{->} 3/2 transition}
\label{sec:iso12iso32}

The full electroweak current can be cast as
\begin{equation}
\myvec{J}^\mu=\myvec{V}^\mu-\myvec{A}^\mu\eKomma
\end{equation}
with
\begin{align}
\myvec{V}^\mu & = - \sqrt{\frac32} \, \mathcal{V}^\mu \, \myvec{T}^\dagger \eKomma \\
\myvec{A}^\mu & = - \sqrt{\frac32} \, \mathcal{A}^\mu \, \myvec{T}^\dagger \eKomma
\end{align}
where $\myvec{T}^\dagger$ is the 1/2 $\to$ 3/2 transition operator
\begin{align}
  T^\dagger_{\pm1} = \mp \frac{T^\dagger_1 \pm \ii T^\dagger_2}{\sqrt{2}}\eKomma \quad  T^\dagger_0=T^\dagger_3\eKomma \\
   \matrixelement{ \frac32 M}{ T^\dagger_\lambda} { \frac12 m}= \clebsch{\frac12 m 1 \lambda}{\frac32 M} \ePunkt
\end{align}
$V^\mu$ and $A^\mu$ are given in \refeq{eq:vectorspinhalfcurrent} and \eqref{eq:axialspinhalfcurrent} for spin 1/2 and in \refeq{eq:vectorspinthreehalfcurrent} and \eqref{eq:axialspinthreehalfcurrent} for spin 3/2 resonances in terms of the corresponding form factors.

The transition current between $I=1/2$ and $I=3/2$ has to be purely isovector, so that
\begin{equation}
  J_{\subEM}^\mu=V_3^\mu=V_0^\mu=- \sqrt{\frac32} \, \mathcal{V}^\mu T^\dagger_0\eKomma
\end{equation}
which yields for the electromagnetic transition matrix element
\begin{align}
\matrixelement{\Delta^+}{J_{\subEM}^\mu}{p}&=- \sqrt{\frac32} \, \mathcal{V}^\mu \clebsch{\frac12 \; \frac12 \; 1 \; 0}{\frac32 \; \frac12} = -\mathcal{V}^\mu \eKomma \\
\matrixelement{\Delta^0}{J_{\subEM}^\mu}{n}&=- \sqrt{\frac32} \, \mathcal{V}^\mu \clebsch{\frac12 \;-\frac12 \;1 \;0}{\frac32 \;-\frac12}= -\mathcal{V}^\mu \ePunkt
\end{align}
The prefactor of $\sqrt{\frac32}$ is chosen such that the electromagnetic matrix element carries no isospin factor. The minus-sign in front is convention --- it is consistent with what is commonly found in the literature \cite{Schreiner:1973mj}. Note that we do not include this minus sign in the current, but instead we include it in the isospin 3/2 form factors (cf.~\reftab{tab:ff}).

For the vector part of the weak charged current, we obtain
\begin{equation}
  V_{\subCC}^\mu=V^{\mu}_1 + \ii V^{\mu}_2 = -\sqrt{2} \left(- \sqrt{\frac32} \, \mathcal{V}^\mu  \right) T^\dagger_{+1}= \sqrt{3}  \, \mathcal{V}^\mu  T^\dagger_{+1}\eKomma
\end{equation}
and therefore
\begin{align}
\matrixelement{\Delta^{++}}{V_{\subCC}^\mu}{p}&= \sqrt{3} \, \mathcal{V}^\mu \clebsch{\frac12 \; \frac12 \; 1 \; 1}{\frac32 \; \frac32}= \sqrt{3} \mathcal{V}^\mu\eKomma \\
\matrixelement{\Delta^+}{V_{\subCC}^\mu}{n}&= \sqrt{3} \, \mathcal{V}^\mu \clebsch{\frac12 \;-\frac12 \;1 \;1}{\frac32 \;\frac12} = \mathcal{V}^\mu \ePunkt
\end{align}
This agrees with the standard expression found in the literature.

Analogously for the axial part of the charged current,
\begin{equation}
  A_{\subCC}^\mu=A^{\mu}_{1} + \ii A^{\mu}_{2}=\sqrt{3}  \, \mathcal{A}^\mu  T^\dagger_{+1} \ePunkt
\end{equation}

For neutral current --- again, only the isovector part of \refeq{eq:vectorcurrentNC} and \refeq{eq:axialcurrentNC} is present --- we have
\begin{equation}
  J_\subNC^\mu=(1- 2\,\sinwein) V_3^\mu - A_3^\mu\eKomma
\end{equation}
and thus
\begin{equation}
 \matrixelement{\Delta^+}{J_{\subNC}^\mu}{p}= \matrixelement{\Delta^0}{J_{\subNC}^\mu}{n} = (1-2\, \sinwein) \, \mathcal{V}^\mu - \mathcal{A}^\mu \ePunkt
\end{equation}

\subsection{Isospin relations for the non-resonant background}
\label{sec:isoBG}

In the case of charged current neutrino scattering, we have three pion production channels, namely
\begin{align}
\nu p &\to l^- p \pi^+ \\
\nu n &\to l^- n \pi^+ \\
\nu n &\to l^- p \pi^0\ePunkt
\end{align}
The vector contribution $V^\mu_\subCC$ of the above reactions is related to the
electromagnetic current $J^\mu_{\subEM}$ via~\cite{LeitnerDr}
\begin{align}
\matrixelement{p \pi^+}{V^\mu_{\subCC} }{p} &=  \sqrt{2} \matrixelement{n \pi^0}{J^\mu_{\subEM}}{n} + \matrixelement{p \pi^-}{J^\mu_{\subEM}}{n} \eKomma  \nonumber \\
\matrixelement{n \pi^+}{V^\mu_{\subCC} }{n} &=  \sqrt{2} \matrixelement{p \pi^0}{J^\mu_{\subEM}}{p} - \matrixelement{p \pi^-}{J^\mu_{\subEM}}{n} \eKomma \nonumber \\
\matrixelement{p \pi^0}{V^\mu_{\subCC} }{n} &=  \matrixelement{p \pi^0}{J^\mu_{\subEM}}{p} - \matrixelement{n \pi^0}{J^\mu_{\subEM}}{n} \nonumber \\ &\quad  - \sqrt{2} \matrixelement{p \pi^-}{J^\mu_{\subEM}}{n} \ePunkt
\end{align}


\section{Form factors and helicity amplitudes}
\label{ch:formfactors_helicityamplitudes}

Here we give details on the connection between the electromagnetic resonance form factors and helicity amplitudes which we apply to obtain the form factors from the helicity amplitudes provided by MAID~\cite{MAIDWebsite,Tiator:2006dq,Drechsel:2007if}.

The helicity amplitudes describe the nucleon-resonance transition depending on the polarization of the incoming photon and the spins of the baryons; they can be defined in various ways \cite{Warns:1989ie, Stoler:1993yk, Aznauryan:2008us}.
We use the following notation
\begin{align}
A_{1/2}=&\sqrt{\frac{2\pi \alpha}{k_R}}  \matrixelement{ R,J_z=1/2 }{ \epsilon_\mu^{+} J^{\mu}_\subEM}{ N,J_z=-1/2}  \zeta\eKomma \nonumber \\
A_{3/2}=&\sqrt{\frac{2\pi \alpha}{k_R}}  \matrixelement{ R,J_z=3/2 }{ \epsilon_\mu^{+} J^{\mu}_\subEM}{ N,J_z=1/2} \zeta \eKomma  \label{eq:helicityamplitudes} \\
S_{1/2}=&-\sqrt{\frac{2\pi \alpha}{k_R}} \frac{\absq}{\sqrt{Q^2}} \nonumber \\
& \times \matrixelement{ R,J_z=1/2 }{ \epsilon_\mu^{0} J^{\mu}_\subEM}{ N,J_z=1/2}  \zeta\eKomma  \nonumber
\end{align}
where $k_R=({\Mpr}^2-M_N^2)/2 \Mpr$ and
\begin{equation}
\absq^2=\frac{({\Mpr}^2-M_N^2-Q^2)^2}{4 {\Mpr}^2} +Q^2\ePunkt
\end{equation}
$\Mpr=\sqrt{\ppr^2}$ is the resonance mass; $J^\mu_\subEM$ is the electromagnetic transition current.
The phase $\zeta$ is given by the relative sign between the $\pi N N$ and $\pi N R$ couplings \cite{Warns:1989ie, Aznauryan:2008us}, which we have taken to be $+1$.

The photon polarization vectors are given by
\begin{equation}
\epsilon^\mu_{\pm}=\mp \frac{1}{\sqrt{2}} (0,1,\pm \ii,0) \quad \quad \text{(transverse)}\eKomma
\end{equation}
and for a photon of momentum $q$ moving along the $z$ axis
\begin{equation}
\epsilon^\mu_{0}= \frac{1}{\sqrt{Q^2}} (\absq,0,0,q^0) \quad \quad  \text{(longitudinal)}\ePunkt
\end{equation}

Our definition of the helicity amplitudes is the same as adopted by MAID (cf.~Eq.~(24) in \refcite{Drechsel:2007if}; $\rho=\myvec{q}\cdot\myvec{J}/q^0$ as introduced in Eq.~(5) of \refcite{Drechsel:1992pn}). While $A_{1/2}$ and $A_{3/2}$ are transverse and Lorentz invariant, $S_{1/2}$ is frame dependent. To be consistent with MAID, the resonance rest frame has to be used for the calculation \cite{tiatorprivcomm}.
We use the outcome of the MAID2005 analysis\footnote{See Section 2 of \refcite{Drechsel:2007if} for a history of MAID and the difference between different versions.}. Note that the helicity amplitudes are taken at the Breit-Wigner mass $M_R$, thus we set $\Mpr=M_R$ in the following.

\subsection{Spin 1/2}
 \label{sec:heliFF_1_2}
 For spin $1/2$ resonances, only the $A_{1/2}$ and $S_{1/2}$ amplitudes are present.
\subsubsection{Positive parity}
In the case of positive parity, the electromagnetic current $J_\subEM^{\mu}$ is given by the one defined in \refeq{eq:vectorspinhalfcurrent}.
We obtain
\begin{align}
A_{1/2}^{p,n}&=\sqrt{\frac{2 \pi \alpha}{M_N} \frac{(M_R-M_N)^2+Q^2}{M_R^2-M_N^2}}  \nonumber \\ & \times \left[  \frac{Q^2}{4 M_N^2} F_1^{p,n} +
\frac{M_R+M_N}{2 M_N} F_2^{p,n} \right]\eKomma\label{eq:spin12heliposA12}
\end{align}
and
\begin{align}
S_{1/2}^{p,n}&=-\sqrt{\frac{ \pi \alpha}{M_N} \frac{(M_N+M_R)^2+Q^2}{M_R^2-M_N^2}} \nonumber \\ & \times \frac{(M_R-M_N)^2 +Q^2}{4 M_R M_N} \left[
\frac{M_R+M_N}{2 M_N} F_1^{p,n} - F_2^{p,n} \right]\ePunkt\label{eq:spin12heliposS12}
\end{align}

\subsubsection{Negative parity}
In the case of negative parity, $J_\subEM^{\mu}$ is given in \refeq{eq:vectorspinhalfcurrent} with an additional $\gF$ (cf.~\refeq{eq:spinhalfcurrentnegparity}) and we find
\begin{align}
A_{1/2}^{p,n}&= \sqrt{\frac{2 \pi \alpha}{M_N} \frac{(M_R+M_N)^2+Q^2}{M_R^2-M_N^2}} \nonumber \\ & \times \left[  \frac{Q^2}{4 M_N^2}
F_1^{p,n} + \frac{M_R-M_N}{2 M_N} F_2^{p,n} \right]\eKomma
\end{align}
and
\begin{align}
S_{1/2}^{p,n}&=\sqrt{\frac{ \pi \alpha}{M_N} \frac{(M_N-M_R)^2+Q^2}{M_R^2-M_N^2}} \nonumber \\ & \times \frac{(M_R+M_N)^2 +Q^2}{4 M_R M_N} \left[
\frac{M_R-M_N}{2 M_N} F_1^{p,n} - F_2^{p,n} \right]\ePunkt
\end{align}

\subsection{Spin 3/2}
\label{sec:heliFF_3_2}

\subsubsection{Positive parity}
In the case of positive parity, the electromagnetic current is given in \refeq{eq:vectorspinthreehalfcurrent}, an additional $\gF$ is needed for positive parity.
This yields
\begin{widetext}
\begin{align}
A_{1/2}^{p,n}=\sqrt{\frac{\pi \alpha}{3 M_N}  \frac{(M_R-M_N)^2+Q^2}{M_R^2-M_N^2}}  \left[ \frac{C^{p,n}_3}{M_N} \frac{M_N^2+M_N M_R +Q^2}{M_R} -  \frac{C^{p,n}_4}{M_N^2} \frac{M_R^2-M_N^2 -
Q^2}{2} -  \frac{C^{p,n}_5}{M_N^2} \frac{M_R^2-M_N^2 + Q^2}{2} \right]\eKomma
\end{align}
\begin{align}
A_{3/2}^{p,n}= \sqrt{\frac{\pi \alpha}{M_N}  \frac{(M_R-M_N)^2+Q^2}{M_R^2-M_N^2}}   \left[ \frac{C^{p,n}_3}{M_N} (M_N+M_R) +  \frac{C^{p,n}_4}{M_N^2} \frac{M_R^2-M_N^2 - Q^2}{2} +
\frac{C^{p,n}_5}{M_N^2} \frac{M_R^2-M_N^2 + Q^2}{2} \right]\eKomma
\end{align}
and
\begin{multline}
S_{1/2}^{p,n} =  \sqrt{\frac{\pi \alpha}{6 M_N}  \frac{(M_R-M_N)^2+Q^2}{M_R^2-M_N^2}} \\ \times  \frac{\sqrt{[(M_R-M_N)^2
+Q^2][(M_R+M_N)^2 +Q^2]}}{M_R^2}
 \left[ \frac{C^{p,n}_3}{M_N} M_R +  \frac{C^{p,n}_4}{M_N^2} M_R^2 +  \frac{C^{p,n}_5}{M_N^2}
\frac{M_R^2+M_N^2 + Q^2}{2}  \right]\ePunkt
\end{multline}
\end{widetext}

\subsubsection{Negative parity}
In the case of negative parity, the electromagnetic current is defined in \refeq{eq:vectorspinthreehalfcurrent} and we get
\begin{widetext}
\begin{align}
A_{1/2}^{p,n}=\sqrt{\frac{\pi \alpha}{3 M_N}  \frac{(M_R+M_N)^2+Q^2}{M_R^2-M_N^2}}  \left[ \frac{C^{p,n}_3}{M_N} \frac{M_N^2-M_N M_R +Q^2}{M_R} - \frac{C^{p,n}_4}{M_N^2} \frac{M_R^2-M_N^2 -
Q^2}{2} -  \frac{C^{p,n}_5}{M_N^2} \frac{M_R^2-M_N^2 + Q^2}{2} \right]\eKomma
\end{align}
\begin{align}
A_{3/2}^{p,n}= \sqrt{\frac{\pi \alpha}{M_N}  \frac{(M_R+M_N)^2+Q^2}{M_R^2-M_N^2}}  \left[ \frac{C^{p,n}_3}{M_N} (M_N-M_R) -  \frac{C^{p,n}_4}{M_N^2} \frac{M_R^2-M_N^2 - Q^2}{2} -
\frac{C^{p,n}_5}{M_N^2} \frac{M_R^2-M_N^2 + Q^2}{2} \right] \eKomma
\end{align}
and
\begin{multline}
S_{1/2}^{p,n} =  -\sqrt{\frac{\pi \alpha}{6 M_N}  \frac{(M_R+M_N)^2+Q^2}{M_R^2-M_N^2} } \\ \times  \frac{\sqrt{[(M_R-M_N)^2
+Q^2][(M_R+M_N)^2 +Q^2]}}{M_R^2}   \left[ \frac{C^{p,n}_3}{M_N} M_R +  \frac{C^{p,n}_4}{M_N^2} M_R^2 +  \frac{C^{p,n}_5}{M_N^2}
\frac{M_R^2+M_N^2 + Q^2}{2}  \right]\ePunkt
\end{multline}
\end{widetext}


\section{Derivation of axial couplings}
\label{ch:pcacaxialcoupling}

In this section, we derive the relations between the axial form factors and
determine the axial couplings applying PCAC according to which the divergence of the axial current is proportional
to the pion mass squared. Thus, in the chiral limit the axial current is conserved
if we assume pion pole dominance, i.e., we assume that the pseudoscalar part is dominated by the one-pion exchange process where a pion is
created at the nucleon-resonance vertex and then couples to the lepton pair. Its current is given by
\begin{align}
  A^\mu_\pi =   (n \to R^+ \pi^-)  \times \left( \frac{\ii}{q^2-\mpi^2}  \right) \times \left( - \ii \sqrt{2} f_\pi q^{\mu} \right)\eKomma      \label{eq:pionpolecurrent}
\end{align}
where the pion decay constant is given by $f_{\pi}=93\MeV$. For the hadronic vertex, we need the interaction Lagrangians which are given in the following.

\subsection{Resonance interaction Lagrangians and widths}
\label{sec:nrpilagrangian}

Here we collect the relativistic Lagrangians used for the descriptions of the coupling of baryon resonances to nucleons and pions.

\subsubsection{Resonances with spin 1/2}

The $R_{1/2}N\pi$ coupling is described by the pseudovector Lagrangian
\begin{equation}
  \mathcal{L}_{R_\subhalf N \pi} = \frac{f}{m_\pi}\bar{\Psi}_{R_\subhalf} \; \binom{\gamma^\mu\gamma_5}{\gamma^\mu} \;
  \partial_\mu \myvec{\phi} \cdot \myvec{t} \,\Psi\eKomma
  \label{eq:spinhalflagrangian}
\end{equation}
where the upper (lower) operator holds for positive (negative) parity resonances and $\myvec{t}=\myvec{\tau}$ ($\myvec{t}=\myvec{T^\dagger}$) for $I=1/2$ ($I=3/2$) resonances.

From this Lagrangian we deduce the following vertex factor
\begin{equation}
-\ii \; C_\text{iso} \;\frac{f}{m_\pi} \; \binom{\gamma^\mu\gamma_5}{\gamma^\mu}\ePunkt \label{eq:spinhalfpiNRvertex}
\end{equation}
The isospin factor is $C_\text{iso} = \sqrt{2}$ for $I=1/2$ and $C_\text{iso}=-\sqrt{\frac13}$ for $I=3/2$ resonances, respectively.

The coupling $f$ can be obtained from the $R_\subhalf \to\pi N$ partial decay width according to
\begin{equation}
\Gamma_{R_\subhalf \to\pi N}= \frac{I_R}{4\pi} \left(\frac{f}{m_\pi}\right)^2 \left(M_R \pm M_N\right)^2 \frac{E_N \mp M_N}{M_R} |\myvec{q}_{\mathrm{cm}}|\eKomma
\label{eq:spinhalfwidth}
\end{equation}
where $|\myvec{q}_{\mathrm{cm}}|$ is the momentum of the outgoing pion in the resonance rest frame
\begin{equation}
|\myvec{q}_{\mathrm{cm}}| = \frac{\sqrt{(M_R^2-m_{\pi}^2-M_N^2)^2 - 4 m_{\pi}^2 M_N^2}}{2 M_R}\ePunkt  \label{eq:pionmomentumrestframe}
\end{equation}
$I_R=1$ for isospin 3/2 and $I_R=3$ for isospin 1/2 resonances, and $E_N$ is the energy of the outgoing nucleon in the resonance rest frame given by
\begin{equation}
  E_N=\frac{M_R^2+M_N^2-m_{\pi}^2}{2 M_R}\ePunkt
\end{equation}

The couplings $f/\mpi$ are determined via \refeq{eq:spinhalfwidth} using the resonance properties given in \reftab{tab:included_resonances}.
To be consistent with the choice of $\zeta$ in \refeq{eq:helicityamplitudes}, $f/\mpi$ has to be positive \cite{Warns:1989ie,Aznauryan:2008us}.

\subsubsection{Resonances with spin 3/2}

The $R_{3/2}N\pi$ coupling is described by the Lagrangian
\begin{equation}
\mathcal{L}_{R_\subthreehalf N \pi} = \frac{f}{m_\pi} \bar\Psi_\mu \; \binom{\unitfour}{\gamma_5} \;
\partial^\mu \myvec{\phi} \cdot \myvec{t}\, \Psi \eKomma
\label{eq:spinthreehalflagrangian}
\end{equation}
where $\Psi_\mu$ is a Rarita-Schwinger $J^\pi = 3/2^+$ field. The upper (lower) operator holds for positive (negative) parity resonances and and $\myvec{t}=\myvec{\tau}$ ($\myvec{t}=\myvec{T^\dagger}$) for $I=1/2$ ($I=3/2$) resonances.

From this Lagrangian we derive for the vertex factor
\begin{equation}
-\ii \; C_\text{iso} \; \frac{f}{m_\pi} \; \binom{\unitfour}{\gamma_5}\eKomma  \label{eq:spinthreehalfpiNRvertex}
\end{equation}
with the isospin factor $C_\text{iso} = \sqrt{2}$ for $I=1/2$ and $C_\text{iso}=-\sqrt{\frac13}$ for $I=3/2$ resonances.

The decay width needed to extract the coupling $f$  is --- for resonances with $J^P=\frac32^\pm$ --- given by
\begin{equation}
\Gamma_{R_\subthreehalf \to\pi N}= \frac{I_R}{12\pi} \left(\frac{f}{m_\pi}\right)^2 \frac{E_N \pm M_N}{M_R} |\myvec{q}_{\mathrm{cm}}|^3\eKomma
\label{eq:spinthreehalfwidth}
\end{equation}
with $|\myvec{q}_{\mathrm{cm}}|$, $E_N$ and $I_R$ as above.

Using the resonance properties given in \reftab{tab:included_resonances}, we obtain the couplings $f/\mpi$.

\subsection{PCAC relations}

\subsubsection{Resonances with spin 1/2}
\label{sec:spinhalfpcac}

We begin with deriving the relation between $F_A$ and $F_P$. Starting with \refeq{eq:axialspinhalfcurrent} (multiplied with $\gF$ for negative parity), we obtain
\begin{equation}
  \partial_{\mu} A^{\mu}_{1/2\pm} = \ii \bar{u}(\ppr) \left[ F_A (M_R\pm M_N) + \frac{F_P}{M_N} q^2 \right]  \,\binom{\gF}{\unitfour}\, u(p) \label{eq:divaxialspinhalfcurrent}
\end{equation}
where the upper (lower) operator holds for positive (negative) parity.
PCAC requires that
\begin{equation}
  \partial_{\mu} A^{\mu}_{1/2 \pm} \longrightarrow 0
\end{equation}
in the chiral limit and, with the extrapolation to non-zero pion masses, we find
\begin{equation}
  F_P(Q^2)=  \frac{(M_R \pm M_N)M_N}{Q^2+m_{\pi}^2}\; F_A(Q^2)\eKomma
\end{equation}
where always the upper sign is for parity $+$ and the lower one for parity $-$.

To determine the coupling $F_A(0)$, we calculate the pion pole contribution to the current and compare that to the axial transition current.
Starting from \refeq{eq:pionpolecurrent} and using \refeq{eq:spinhalfpiNRvertex} we find
\begin{equation}
  A^\mu_\pi  =   - \bar{u}(\ppr)  \; C_\text{iso} \;\frac{f}{m_\pi} \; (M_R\pm M_N)  \; \binom{ \gamma_5 }{\unitfour} \;  u(p)  \frac{ \sqrt{2} f_\pi q^{\mu}}{q^2-\mpi^2}\ePunkt
\end{equation}

By comparing $\partial_{\mu} A^{\mu}_\pi$ with the divergence of the axial current, given in \refeq{eq:divaxialspinhalfcurrent}, we obtain
in the limit of $q^2\to 0$
\begin{equation}
  F_A (0)  = - C_\text{iso} \; \sqrt{2} f_\pi \frac{f}{m_\pi}   \label{eq:spin12_axialcoupling}
\end{equation}

With this procedure we calculate the axial couplings $F_A(0)$ as summarized in \reftab{tab:included_resonances}.

\subsubsection{Resonances with spin 3/2}
\label{sec:spinthreehalfpcac}

We first derive the relation between $C_5^A$ and $C_6^A$. Starting with \refeq{eq:axialspinthreehalfcurrent} (multiplied with $\gF$ for positive parity), we obtain
\begin{align}
  \partial_{\mu} A^{\mu}_{3/2\pm}= \ii \bar{\psi}_{\alpha}(\ppr) q^{\alpha} \left[{C_5^A} + \frac{C_6^A}{M_N^2} q^2 \right] \,\binom{\unitfour}{\gamma_{5}} \, u(p)\eKomma \label{eq:divaxialspinthreehalfcurrent}
\end{align}
where the upper (lower) operator holds for positive (negative) parity.
Assuming PCAC we can relate the form factors (same for both parity states)
\begin{equation}
  C_6^A (Q^2)= \frac{M_N^2}{Q^2+m_{\pi}^2}\; C_5^A(Q^2)\ePunkt
\end{equation}

Analogous to the spin 1/2 case, we derive $C_5^A(0)$ by starting from \refeq{eq:pionpolecurrent} and using \refeq{eq:spinthreehalfpiNRvertex}
\begin{equation}
  A^\mu_\pi  =   - \bar{\psi}_\alpha(\ppr) \; C_\text{iso} \; \frac{f}{m_\pi} \; \binom{\unitfour}{\gamma_5} \; q^\alpha \; u(p)  \frac{ \sqrt{2} f_\pi q^{\mu}}{q^2-\mpi^2}\ePunkt
\end{equation}

Comparing $\partial_{\mu} A^\mu_\pi$ to $\partial_{\mu} A^{\mu}_{3/2\pm}$ (cf.~\refeq{eq:divaxialspinthreehalfcurrent})
yields in the limit $q^2\to 0$
\begin{equation}
C_5^A (0) = - C_\text{iso} \; \sqrt{2} f_\pi \; \frac{f}{m_\pi}\ePunkt
\end{equation}

The final results for the axial couplings $C_5^A(0)$ are given in \reftab{tab:included_resonances}

\subsubsection{Resonances with spin \texorpdfstring{$>$}{>} 3/2}
\label{sec:spinmorethanthreehalfpcac}

We make again the assumption that resonances with spin $>$ 3/2 are described with the corresponding spin 3/2 formula given in the previous sections. The obtained couplings are summarized in \reftab{tab:included_resonances}.


\end{document}